\documentclass[10pt,conference]{IEEEtran}
\usepackage{cite}
\usepackage{amsmath,amssymb,amsfonts}
\usepackage{graphicx}
\usepackage{textcomp}
\usepackage{xcolor}
\usepackage[hyphens]{url}
\usepackage[pdfstartview=XYZ,
bookmarks=true,
colorlinks=true,
linkcolor=blue,
urlcolor=blue,
citecolor=blue,
pdftex,
bookmarks=true,
linktocpage=true, 
hyperindex=true]{hyperref}
\usepackage{orcidlink}
\usepackage{color}
\usepackage{listings}
\usepackage[capitalise]{cleveref}
\usepackage{multirow}
\usepackage{algorithm}
\usepackage{algpseudocode}
\usepackage{minted,xcolor}
\usepackage{marginnote}
\definecolor{mintbg}{gray}{0.95}
\setminted{bgcolor=mintbg, frame=lines, rulecolor=\color{gray!50}}
\setminted[magma.py:Magma -x]{mathescape=true} 

\usepackage{amsmath}

\usepackage{amssymb,amsfonts}
\usepackage{algorithm}
\usepackage{algpseudocode}
\usepackage{mathtools}
\usepackage{graphicx}
\usepackage{textcomp}
\usepackage[dvipsnames]{xcolor}
\usepackage[most]{tcolorbox}
\usepackage{fancyhdr}
\usepackage{hyperref}
\usepackage[capitalise]{cleveref}
\usepackage{multirow}
\usepackage{setspace}
\usepackage{caption}
\usepackage{subcaption}
\usepackage{rotating}
\usepackage{comment}
\usepackage{dblfloatfix}
\usepackage{wrapfig}
\usepackage[absolute,overlay]{textpos}
\usepackage{framed}
\def\BibTeX{{\rm B\kern-.05em{\sc i\kern-.025em b}\kern-.08em
    T\kern-.1667em\lower.7ex\hbox{E}\kern-.125emX}}

\usepackage[T1]{fontenc}
\usepackage{listings}

\newcommand{\hpcayear}{2026}

\newcommand{\hpcasubmissionnumber}{10}

\title{Plutarch: Toward Scalable Operational Parallelism on Racetrack-Shaped Trapped-Ion Processors}

\def\hpcacameraready{} 

\newcommand\hpcaauthors{
Enhyeok Jang$^\dagger$,
Hyungseok Kim$^\dagger$,
Yongju Lee$^\dagger$,
Jaewon Kwon$^\dagger$,
Yipeng Huang$^\ddagger$,
and 
Won Woo Ro$^\dagger$
}
\newcommand\hpcaaffiliation{
\textit{
$^\dagger$Yonsei University, Seoul, South Korea
$^\ddagger$Rutgers University–New Brunswick, Piscataway, New Jersey, United States
}
}
\newcommand\hpcaemail{
\textit{Email: \{enhyeok.jang, hyungseok.kim, yongju.lee, jaewon.kwon, wro\}@yonsei.ac.kr, yipeng.huang@rutgers.edu}
}

\author{
  \ifdefined\hpcacameraready
    \IEEEauthorblockN{\hpcaauthors{}}
      \IEEEauthorblockA{
        \hpcaaffiliation{} \\
        \hpcaemail{}
      }
  \else
    \IEEEauthorblockN{\normalsize{HPCA \hpcayear{} Submission
      \textbf{\#\hpcasubmissionnumber{}}} \\
      \IEEEauthorblockA{
        Confidential Draft \\
        Do NOT Distribute!!
      }
    }
  \fi 
}

\fancypagestyle{camerareadyfirstpage}{%
  \fancyhead{}
  
  \fancyfoot[C]{}
}
\usepackage{marginnote}
\usepackage{tikz}
\usepackage[absolute,overlay]{textpos}

\begin{document}

\maketitle

\ifdefined\hpcacameraready 
  \thispagestyle{camerareadyfirstpage}
  \pagestyle{empty}
\else
  \thispagestyle{plain}
  \pagestyle{plain}
\fi

\newcommand{\hpcaheight}{0mm}
\ifdefined\eaopen
\renewcommand{\hpcaheight}{12mm}
\fi

\begin{abstract}

A recent advancement in quantum computing shows a quantum advantage of certified randomness on the racetrack processor. 
This work investigates the execution efficiency of this architecture for general-purpose programs. 
We first explore the impact of increasing zones on runtime efficiency. 
Counterintuitively, our evaluations using variational programs reveal that expanding zones may degrade runtime performance under the existing scheduling policy. 
This degradation may be attributed to the increase in track length, which increases ion circulation overhead, offsetting the benefits of enhanced parallelism.
To mitigate this, the proposed \textit{Plutarch} exploits 3 strategies:
(i) unitary decomposition and translation to maximize zone utilization, 
(ii) prioritizing the execution of nearby gates over ion circulation, and
(iii) implementing shortcuts to provide the alternative path.

\end{abstract}

\section{Introduction} \label{intro}

Quantum computing shows the potential to address computational challenges that are intractable for classical systems \cite{ayanzadeh2024skipper, ludmir2024modeling, kundu2025adversarial}. 
Today, trapped ions are one of the promising qubit candidates \cite{wu2021tilt, wu2024boss, maksymov2022detecting, wang2021single}.
Recently, a demonstration of certified randomness \cite{liu2025certificate} with the H2 \cite{QuantinuumH2Datasheet} shows their capability to provide a quantum advantage \cite{decross2024computational}.
This architecture features a racetrack-shaped electrode as shown in \cref{f1}. 
In the racetrack architecture, the program runs in a Rolodex fashion \cite{decross2024computational}, where each ion cycles the track and stops in the operating zone to perform gate operations or ion reordering instructions.

The racetrack-shaped electrode could be viewed as a compromise between the one-dimensional architecture \cite{dasu2025breaking} and a large 2-dimensional grid architecture \cite{malinowski2023wire}.
In a racetrack, ions circulate along a loop so that some ions can undergo operations in the bottom zones while others can be rearranged in the top zones, providing better flexibility than the H1 \cite{dasu2025breaking}.
Ultimately, a 2D grid module would be required, but large-scale electrode technologies would still require several years of development.

This research is interested in the runtime efficiency of the racetrack architecture for general-purpose programs.
Recently popular applications, such as variational algorithms \cite{ueno2024c3, liu2025hatt, liu2024fermihedral, li2022paulihedral, kim2024distribution}, have execution characteristics different from randomized circuits.
These programs often contain deep two-qubit operations whose computational dependencies restrict the fully parallel execution.
Grouping such dependent two-qubit gates and executing them as quickly as possible, rather than circulating the ions for every gate layer, could be more efficient in terms of runtime.
The most intuitive solution is to scale up the hardware itself.
We can expect that increasing the number of zones reduces the program runtime by increasing the number of gates that could be executed in a single cycle.

However, our experiments reveal a counterintuitive finding: simply increasing the number of zones could degrade runtime performance, under existing rolodex-style execution policies (as discussed in the \cref{tradeoff}). 
This degradation may be attributed to the physical inter-zone interval in the track, which should be ensured to prevent an excessive density of phonon-mode frequencies \cite{wu2024boss, chen2021quantum}.
Additional gate operational zones may proportionally extend the track length, thereby increasing the ion shuttling overhead. 
As the electrode lengthens, the time required to move ions to the appropriate operational zones grows, potentially negating their enhanced parallelism benefit.

To address this, we propose \textit{Plutarch} (\underline{P}rioritized schedu-\underline{L}ing \underline{U}nder a\underline{T}om tr\underline{A}nsport fo\underline{R} ra\underline{C}etrack arc\underline{H}itecture).
\textit{Plutarch} exploits 3 strategies.
First, \textit{Plutarch} revisits whether the conventional wisdom of two-qubit gate placements within programs for other competitive architectures (e.g., superconducting \cite{das2022lilliput, ravi2023better, das2022afs, vittal2023astrea, alavisamani2024promatch, kim2025qr}) is also efficient on the racetrack.
For this, we consider the case of the Z-phase gadget \cite{Cowtan_2020}, a representative sub-routine for the variational programs.
Our evaluations demonstrate that only after using an improved compile strategy that maximizes parallelism can we saturate zone utilization in the current specification (as discussed in the \cref{saturate}).
Second, the proposed scheduling minimizes ion circulation by prioritizing the local execution of 2-qubit gates and their adjacent 1-qubit gates over the ion circulation, as discussed in the \cref{propsched}. 
In doing so, the proposed scheduling could suppress excessive ion circulation and prevent the performance degradation, even if the electrode is further expanded.
The third strategy introduces shortcuts to allow zone-independent ion alignment and ensure the execution performance of smaller programs.
This can enable qubit re-indexing without a reordering zone (as discussed in \cref{f13}).

The evaluations in \cref{results} quantify how \textit{Plutarch} improves the racetrack processor's performance across multiple regimes.
As shown in \cref{specific}, \textit{Plutarch} could reduce the execution time by 71\% for 32-qubit variational workloads.
Additionally, when mapped to the execution budget of the real H2-1 processor, the estimated end-to-end QAOA (quantum approximate optimization algorithm \cite{marwaha2021local, hashim2022optimized, jang2024recompiling}) training time could drop from 41.38 hours under Rolodex to 14.07 hours under \textit{Plutarch}.
As shown in \cref{fidelity}, the fidelity analyses show that the \textit{Plutarch} could achieve 19.73\% lower infidelity than Rolodex and 19.83\% lower than H1-like execution across deep variational programs of up to 256 QAOA layers or VQE (variational quantum eigensolver \cite{ueno2024sfq}) Pauli strings.
As shown in \cref{ft-programs}, \cref{f12} shows runtime reductions of 32\% by \textit{Plutarch} for fault-tolerant workloads based on the [[7,1,3]] Steane code \cite{ryan2021realization, ryananderson2022, paetznick2024, bluvstein2023, metodi2005quantum}.
As shown in \cref{modified}, introducing a few shortcuts could improve runtime by up to 53\% for QAOA and 41\% for magic-state distillation compared to a \texttt{No shortcut} track.
Finally, in the large-scale quantum Reed-Muller code \cite{barg2025geometric}'s encoding study of \cref{shortcut2}, the proposed non-uniform \textit{Halved} layout reduces shuttling cost by 48.99\% relative to an evenly spaced layout, 81.13\% relative to the \texttt{No shortcut} case, and 86.38\% relative to H1-like run, and changes the shuttling scaling for the encoding process from quadratic to linear in the number of qubits, suggesting a viable path for future large-scale QCCD (Quantum Charge-Coupled Device) architectures.

\section{Background and Motivation} \label{background}

\subsection{Trapped-Ion Quantum Processor with Racetrack Electrode}

\begin{figure} [h] 
  \centerline {
  \includegraphics [width=1\columnwidth] {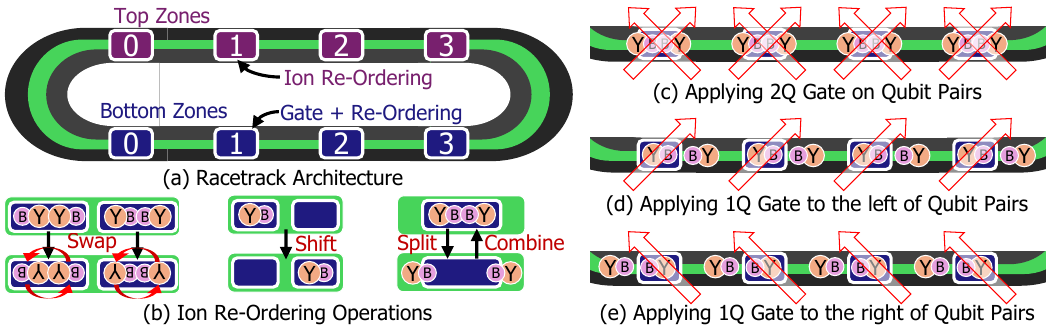} }
  \caption {
    A racetrack architecture \cite{decross2024computational}.
    (a) Top zones (purple), which perform ion reordering operations but do not execute gate operations, and bottom zones (blue), capable of both gate operations and ion reordering.
    Each of the top and bottom regions consists of 4 gate zones. 
    (b) Ion reordering operations (Swap, Shift, Split, and Combine).
    (c) 2-qubit gate operation, which is possible only when ions are arranged in a combined structure of Yb-Ba-Ba-Yb. 
    (d) Single-qubit gate operation to the left-side qubits of each pair after splitting the ions' crystal structures. 
    (e) Single-qubit gate operation to the right-side qubits of each pair after splitting the ions' crystal structures.
  } 
  \label{f1} 
\end{figure}

A racetrack architecture is explained in the \cref{f1}. 
A single qubit is composed of 1 ytterbium ion ($^{171}$Yb$^+$) and 1 barium ion ($^{138}$Ba$^+$), where each $^{171}$Yb$^+$ performs gate operations, and each $^{138}$Ba$^+$ serves as a coolant its paired $^{171}$Yb$^+$ \cite{Cowtan_2020}. 

\textbf{Program Execution via Circulating Ions}: 
The key distinction of racetrack architecture compared to other architectures is its utilization of circulating ions for operations. 
For each circuit depth, up to four homogeneous gate operations can be processed in parallel.
Qubits cycle in a \textit{Rolodex} style, continuously circulating until all gates in the current layer's execution are completed \cite{decross2024computational}. 
After processing the operations in the current gate-layer, the automated compiler \cite{decross2024computational} rearranges the ions using upper zones to handle subsequent layer operations.

\textbf{Native gate} \cite{QuantinuumH2Datasheet}:
The native gate set of the racetrack architecture includes single-qubit rotation gates:
\(U_{1q}(\theta, \phi) = e^{-i\frac{\theta}{2}(\cos(\phi)\hat{X}+\sin(\phi)\hat{Y})}\) and \(R_z(\lambda) = e^{-i\frac{\lambda}{2}\hat{Z}}\); 
fully entanged ZZ interation gate such as \(ZZ() = e^{-i\frac{\pi}{4}\hat{Z}\otimes\hat{Z}}\); 
parameterized (partially) entangled ZZ interation gate \(RZZ(\theta) = e^{-i\frac{\theta}{2}\hat{Z}\otimes\hat{Z}}\); 
and general SU(4) entanglers \(Rxxyyzz(\alpha, \beta, \gamma) = e^{-i\frac{1}{2}(\alpha \hat{X}\otimes\hat{X} + \beta \hat{Y}\otimes\hat{Y} + \gamma \hat{Z}\otimes\hat{Z})}\). 
For example, a single Hadamard gate can be implemented as \(U_{1q}(\pi/2, -\pi/2) \cdot \, R_z(\pi)\). 
A single CX (Controlled-NOT) gate can be implemented as \(CX^{(c,t)} = U_{1q}^{(t)}(-\cfrac{\pi}{2}, \cfrac{\pi}{2}) \,\cdot ZZ() \,\cdot R_z^{(c)}(-\cfrac{\pi}{2}) \, \cdot U_{1q}^{(t)}(\cfrac{\pi}{2}, \pi) \, \cdot R_z^{(t)}(-\cfrac{\pi}{2})\) \cite{moses2023race}.

\subsection{Challenge and Opportunity on the Racetrack Architecture} \label{challenge}

Despite the advantages of the racetrack, the restriction to executing only four simultaneous gates may pose challenges when running high-parallel programs. 
Other architectures, such as superconducting \cite{acharya2024quantum, mckinney2023co, liu2022not, kobori2025lsqca, das2023imitation, ueno2022qulatis, kim2024fault} or neutral atom \cite{evered2023high, viszlai2025interleaved, lin2025reuse} qubit-based platforms, can execute an identical gate across all qubits in a single-depth cycle.
Conversely, the racetrack architecture might face increased circuit depth complexity as program sizes expand.
Applying a 1-qubit native gate to each of 32 qubits takes one cycle for superconducting or neutral atom-based processors, but eight cycles and the extra ions' shuttling time are required for the racetrack architecture.
This motivates investigating increasing gate zones on the racetrack to enhance parallelism.

However, naively adding more gate zones densely to the current racetrack structure may lead to decreased two-qubit gate fidelity due to the excessive clustering of phonon mode frequencies \cite{wu2024boss, chen2021quantum}. 
To maintain high accuracy of the program execution, gate operation zones should be spaced sufficiently far apart, inevitably lengthening the overall track.
Due to this, the standard rolodex-style run, which requires all ions to circulate around the entire track for each gate-layer, may not yield proportional execution efficiency improvements according to the track's expansion.
This is because extra zones to achieve greater gate-level parallelism increase one-lap time.

Despite these hardware challenges, it is still worthwhile exploring execution scenarios when the racetrack architecture scales very large.
Current H2 processors can prepare up to 56 qubits \cite{moses2023race}, exceeding the scale simulated by existing classical supercomputers, but perhaps not in fault-tolerant algorithmic scenarios.
For example, applying the 7-qubit color code allows H2 processors to prepare up to eight logical qubits, which can be easily simulated by classical computing resources.
This work will discuss various hardware and software strategies to improve execution efficiency for achieving a scalable design.

\begin{figure} [h] 
  \centerline {
  \includegraphics [width=0.89\columnwidth] {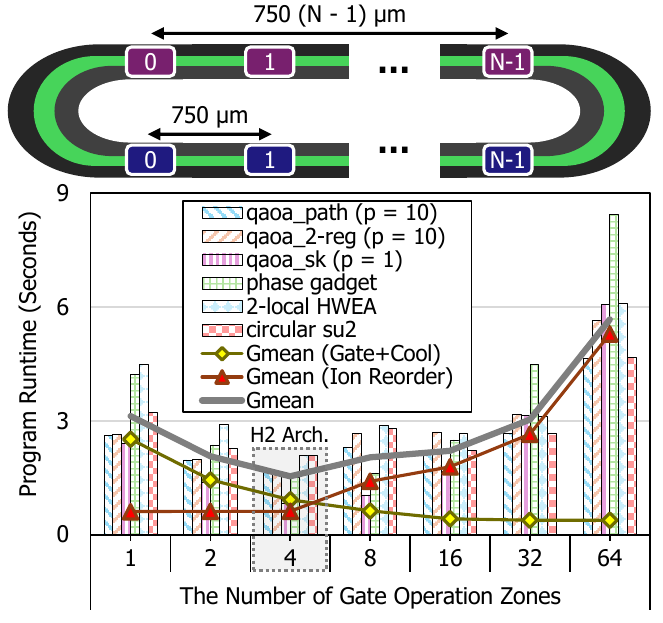} }
  \caption {
    Program runtime analysis by the rolodex scheduling \cite{decross2024computational} according to the number of gate operation zones in the racetrack architecture using 32-qubit QAOA (with target graphs of path-shaped \cite{tannu2019ensemble}, 2-regular \cite{marwaha2021local}, and Sherrington-Kirkpatrick model \cite{farhi2022quantum}) and VQE (phase gadget \cite{Cowtan_2020}, 2-local HWEA \cite{li2024case}, and circular-shaped efficient SU2 \cite{wang2023prepare}) programs. 
    The current H2 processor \cite{QuantinuumH2Datasheet, moses2023race} utilizes 4 operation zones. 
  } 
  \label{f2} 
\end{figure}

\subsection{Expansion of Operation Zones and its Trade-off Analysis} \label{tradeoff}

\cref{f2} shows that various variational programs could run most efficiently under the rolodex policy for four zones.
This is because increasing zones can introduce a contradictory effect: improving gate-level parallelism reduces gate application time, while their extended electrode structure potentially increases shuttling time.
For example, as the circulation time of ions is no longer hidden in computation time, the ion-reordering overhead increases when the gate zones are more than 4.
Note that 16 parallel two-qubit gates are processed in two batches when the number of zones is 8.
It takes about 6 $ms$ to process these batched gates, whereas the 1-lap time of ions for this track takes about twice as much, 12.4 $ms$.
This means that the wait time is required from the end of processing the current gate-layer batch until the next batch is fetched, and the circulation time may no longer be hidden by gate applications.

When there are fewer than 4 zones, the overall runtime can be dominated by the gate applications, as their limited parallelism requires more cycles to run parallel homogeneous gates. 
As the zone count increases from 1 to 32, the gate application time could steadily decrease as more simultaneous gates are allowed. 
This gate operation time could no longer be improved beyond 32 gate zones, since the maximum achievable parallelism for 32-qubit programs has been reached.

The rolodex-style run (which requires ion's circulation for every batch) incurs a near-proportionally longer travel time, as the linear track length would expand with each new gate zone, as discussed in \cref{challenge}. 
Although 32 zones yield the shortest gate and cooling times, the overall runtime (3.04 seconds) is nearly twice that of the 4-zone scenario (1.54 seconds). 
The trade-off between gate-level parallelism gains and increased track length indicates that adding more zones may not provide remarkable benefits.
Thus, while increasing the number of zones is inevitable to improve gate-level parallelism in racetrack architectures, developing scheduling techniques that suppress ion circulation would be essential to achieve scalable performance gains as the number of zones increases.

\begin{tcolorbox}[colback=green!10, colframe=green!0, boxrule=0pt, left=0.1mm, right=0.1mm, top=0.1mm, bottom=0.1mm]
\textbf{Section Summary:}
Adding zones in the racetrack architecture to improve parallelism could increase shuttling time overhead for ions to circulate the entire racetrack, which may be larger than the time saved by parallelized gate executions.
\end{tcolorbox}

\section{Efficient Hamiltonain Compile for Racetrack} \label{decomp}

The first step in architecture-specific execution optimization could start from decomposing the unitaries derived from the algorithm to suit target hardware characteristics.
This section proposes an efficient design for the 2-qubit gate chain tailored to the racetrack trapped-ion architecture and explains why this structure is efficient from the execution scheduling perspective.  
For the explanation, we consider the case study of the fermionic system simulation \cite{liu2024fermihedral, li2022paulihedral, jin2024tetris, liu2025hatt}, which is a representative application of near-term quantum computing.  
The Z-phase gadget is an iterative subroutine in the simulation kernel.
Once an efficient 2-qubit gate decomposition from a high-level Hamiltonian for the racetrack architecture is determined, generalizing it to general programs is straightforward.

\subsection{Efficient 2Q Gates' Placement for Racetrack Architectures} \label{2QPlace}

\begin{figure} [h] 
  \centerline {
  \includegraphics [width=1\columnwidth] {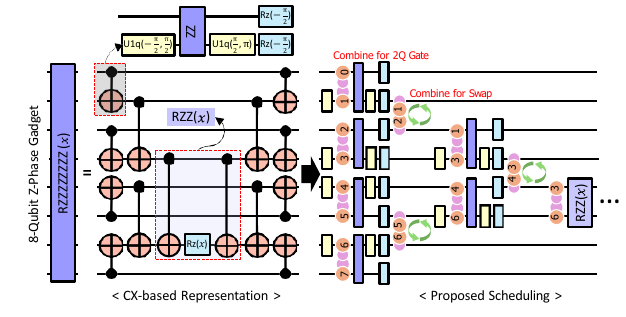} }
  \caption {
    An example of the efficient gate decomposition and translation for an 8-qubit Z-phase gadget. 
    As shown in the left half of the picture, a CX chain structure whose circuit depth expands logarithmically \cite{cruz2019efficient, jang2025qubit} according to the number of qubits is adopted to maximize the utilization of zones.
    In order to minimize the ion reordering overhead, the inner chain is gradually formed at the center of the chain.
    In the right half of the picture, most CX gates are translated to the ion's native gates.
    The innermost chain of two CXs and one RZ can be directly implemented as a single native ZZ-interaction operation. 
    Although the translated representation omits the right-half operations, they can be executed by applying the gates in the reverse order, analogous to the left-half gate chain. 
  } 
  \label{f3} 
\end{figure}

An example of efficient decomposition and translation of the Z-phase gadget \cite{Cowtan_2020} proposed in this work is shown in \cref{f3}.
The principle of efficient 2-qubit gate placement from the high-level Hamiltonian for the racetrack architecture could be explained by three primary design considerations as follows.

\subsubsection{Maximizing utilization of gate zones}
In the racetrack architecture, a dominant portion of the total runtime is attributable to gate application time and their subsequent cooling process \cite{decross2024computational}. 
Therefore, it is important to utilize all four given gate zones as much as possible during runtime.
This necessity drives the racetrack architecture to prefer a high-parallel chain structure compared to other quantum platforms. 
For example, superconducting qubit-based processors with heavy-hex topology benefit from sparse-connectivity-friendly chain configurations such as V-shaped or ladder-shaped chains \cite{Cowtan_2020} due to routing efficiency \cite{jin2024tetris}. 
Meanwhile, the zoned neutral atom-based processor may employ a fountain-shaped chain that minimizes trap-transfer overhead \cite{jang2025qubit}. 
However, these configurations generally allow no more than two native gates to be executed simultaneously per layer, resulting in the under-utilization of gate zones in the racetrack architecture. 
To maximize gate zone utilization for the racetrack architecture, arranging 2-qubit gate chains in parallel can be advantageous.

\subsubsection{Minimizing alignment overhead of ions}
Among parallel 2-qubit gate-chain structures whose depth scales logarithmically with the number of qubits, it is beneficial to the racetrack architecture for alternating the control-target direction of each CX gate rather than maintaining a uniform direction, as shown in the \cref{f3}. 
This allows the re-combination for the 2-qubit execution of Yb-Ba-Ba-Yb without the need for extra ion-pairs' rotations, thereby reducing reordering overhead.
An additional advantage of alternating control-target directions is that the internal CXs gradually form towards the central region of the overall ions' chain. 
This allows qubits from both sides of the ion sequence to approach each other by moving similar distances for re-combination from each other, reducing the shuttling overhead compared to the CX gate structures of uniformly directed control-target.
The method for generating these proposed Z-phase gadgets is detailed in the \cref{appendixa}.

\subsubsection{Exploiting native gate operations of trapped-ion qubits}
The racetrack architecture natively supports arbitrary-angle ZZ interaction gates. 
The RZZ operation is commonly included inside all Z-phase gadget operations, which can be translated directly into native gates instead of being decomposed into 2 CX and RZ, saving the number of gates to run from 11 to 1.

\subsection{Generalization of Efficient Phase Gadget Implementation} \label{appendixa}

This section describes a general method for generating an efficient two-qubit gate arrangement described in \cref{2QPlace} given an arbitrary Z-phase gadget.
Given an ordered set of \(n\) qubits \(Q = \{q_0, q_1, \dots, q_{n-1}\}\) and a rotational parameter \(\alpha\), we define the balanced-tree Z-phase gadget operator \(\Phi_n(\alpha)\) as follows. 
The control and target of the CX gate are applied alternately to facilitate the re-combining ions along the Yb-Ba-Ba-Yb crystal structure for the next layer's two-qubit gates.

First, we recursively define a sequence of active qubit subsets \(Q^{(k)}\), starting from \(Q^{(0)} = Q\). At each recursion step \(k\), given the active set \(Q^{(k)} = \{q_{Q^{(k)}_0}, q_{Q^{(k)}_1}, \dots, q_{Q^{(k)}_{|Q^{(k)}|-1}}\}\), we form pairs of adjacent qubits and apply a layer of parallel CXs. 
The order of the control and target qubits for these CXs alternates at every recursion level: 
when \(k\) divided by 4 is less than or equal to 1 (i.e., 0 or 1), we apply gates of the form \(\text{CX}(q_{Q^{(k)}_{2j+1}}\rightarrow q_{Q^{(k)}_{2j}})\), and when \(k\) divided by 4 is greater than 1 (i.e., 2 or 3), we reverse the direction, applying \(\text{CX}(q_{Q^{(k)}_{2j}}\rightarrow q_{Q^{(k)}_{2j+1}})\), for the each \(j=0,\dots,\lfloor|Q^{(k)}|/2\rfloor - 1\).

After applying gates in the first layer, we redefine the indexing for the active qubit set at the next recursion step \(Q^{(k+1)}\) to consist exclusively of the target qubits from the CX gates at step \(k\). 
This allows the active qubit set for the next recursion to be composed of the previous layer's target qubits at every step. 
If at any step the active set \(Q^{(k)}\) has an odd number of qubits, the unpaired last qubit moves into the active set \(Q^{(k+1)}\). 
We repeat this recursion until only a single active qubit remains. 
This re-indexing can be guaranteed by the following definitions: 
at even recursion steps, the next active set is \(Q^{(k+1)}:= \{q_{Q^{(k)}_{2j}}\mid j=0,\dots,\lfloor|Q^{(k)}|/2\rfloor-1\}\), while at the odd recursion steps, it becomes \(Q^{(k+1)}:= \{q_{Q^{(k)}_{2j+1}}\mid j=0,\dots,\lfloor|Q^{(k)}|/2\rfloor-1\}\).
The left side of the chain \(CX_{T}(Q)\) can be defined by the following product from \(k=0\) to \(k=K-1\):
\[
\prod_{k=0}^{K-1}
\left(
\bigotimes_{j=0}^{\lfloor|Q^{(k)}|/2\rfloor-1}
\begin{cases}
\text{CX}(q_{Q^{(k)}_{2j+1}}\rightarrow q_{Q^{(k)}_{2j}}), & \text{if } k \% 4 <= 1 \\[5pt]
\text{CX}(q_{Q^{(k)}_{2j}}\rightarrow q_{Q^{(k)}_{2j+1}}), & \text{else }
\end{cases}
\right).
\]

Then, we apply the \(R_z(\alpha)\) to the final active qubit (which means the target qubit of the last applied CX): \(I_{n-1}\otimes R_z(\alpha)\). 
This \(R_z(\alpha)\) can be grouped with its two neighboring CXs and rewritten as \(R_{zz}(\alpha)\).
Finally, we apply the mirrored inverse of the alternated balanced-tree CX chain, denoted as \(CX_{T}(Q)^\dagger\).
Combining these steps, the proposed Z-phase gadget can be given by the product: 
$ CX_{T}(Q) \cdot \left(I_{n-1}\otimes R_z(\alpha)\right) \cdot CX_{T}(Q)^\dagger$.

The computational complexity of our phase-gadget generation algorithm is $O(n)$ according to the number of qubits $n$.
This is because the gadget structure consists of $2(n-1)$ CXs and a single $R_z$.
In our evaluation conducted on a system equipped with an Intel Jasper Lake N5095 processor and 32 GB of DDR4 memory, generating and saving a 32-qubit phase gadget required 0.3 ms, while a 1-million-qubit phase gadget was generated and stored in 1.8 seconds.
This shows that the proposed algorithm scales linearly and would remain sufficiently efficient for the large-scale fault-tolerant systems.

\subsection{Execution Time Analysis According to 2Q Chain Structures} \label{saturate}

\begin{figure} [h] 
  \centerline {
  \includegraphics [width=1\columnwidth] {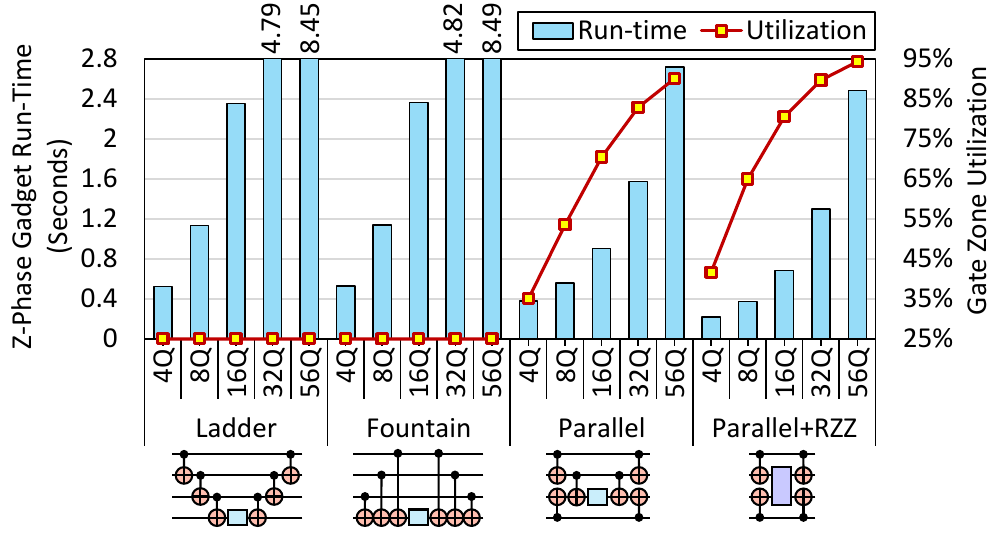} }
  \caption {
    Comparison of runtime and zone utilization for various Z-phase gadgets (Ladder, Fountain, Parallel, and Parallel+RZZ) according to different qubits. 
    Lower runtime and higher zone utilization mean better performance, respectively.
  } 
  \label{f4} 
\end{figure}

Z-phase gadgets can have various (and semantically equivalent) arrangements of two-qubit gates, as they have the flexibility of synthesis \cite{li2022paulihedral}.
This section analyzes the performance according to two-qubit gate arrangements that constitute Z-phase gadgets on the racetrack architecture.
Zone utilization is defined as the percentage of gate zones engaged in gate application or cooling during periods when gate zones are in use. 
For example, if an average of two gate zones is activated throughout the run, the utilization rate is 50\%.  
As illustrated in \cref{f4}, ladder or fountain chain structures consistently exhibit low gate zone utilization of about 25\%, regardless of the number of qubits, because they execute only 1 gate per layer. 

These results indicate that 2-qubit gate placements previously proven efficient on other architectures (such as ladder chains for hexagon-shaped qubit topology \cite{Cowtan_2020} or fountain chains for neutral atom-based device \cite{jang2025qubit}) may not be optimal for the racetrack architecture. 
Compared to ladder and fountain chains, the parallel chain structure improves runtime by over 3.1$\times$ and gate zone utilization by 3.6$\times$, achieving 90.2\% utilization at 56 qubits.  
Furthermore, replacing the innermost chain (which is the most challenging part to exploit parallelism in the phase gadget) with a single RZZ could further enhance performance. 
Relative to the standard parallel chain, this method reduces runtime by an additional 9\% and improves the zone utilization by 5\%, reaching 94.4\% utilization at 56 qubits.
This proposed 2-qubit gate chain structure not only reduces runtime but also provides the benefit of improved fidelity via the reduced coherence error caused by circuit depth reduction.

These evaluation results indicate that low-level programs generated by conventional compilers for fixed qubits or sparse topologies may not be executed optimally on the racetrack architecture.
To run efficiently on the racetrack architecture, the same kind of gates should be placed in parallel as much as possible, the initial qubits should be aligned and mapped so that the gate chain eventually forms toward the center of the ion sequence, and 2-qubit native gates should be considered during the translation process.
These design considerations are not limited to simulation kernels, but can readily extend to applications that require complex structures of two-qubit gates.  

\begin{tcolorbox}[colback=green!10, colframe=green!0, boxrule=0pt, left=0.1mm, right=0.1mm, top=0.1mm, bottom=0.1mm]
\textbf{Section Summary:}
We investigate the execution efficiency of a 2-qubit gate chain for the racetrack architecture using the case study of the Z-phase gadget.
We propose an efficient unitary decomposition from the perspective of high utilization of zones, low shuttling, and exploiting native operations.
\end{tcolorbox}

\section{Prioritize Opportunities to Run Nearby Gates} \label{lewis2}

In the racetrack architecture, the rolodex scheduling \cite{decross2024computational, moses2023race} firstly groups gates simultaneously runnable within each layer into batches and executes these batches sequentially. 
While the rolodex scheduling has a straightforward and effective greedy algorithm, it could not consider concurrently executable sub-blocks of gates between different layers and thus might not be optimal from a global scheduling perspective. 
We observe that single-qubit gates adjacent to a two-qubit gate can be executed in place, which do not have to be executed only after going around the racetrack.
By identifying these sub-blocks and then pre-emptively running them, the overhead of ions' circulation per gate-layer could be reduced.
Furthermore, when the number of operation zones is sufficient, executing the program in a one-dimensional fashion (without circulating ions on the racetrack) can offer better runtime performances.

The proposed scheduling works as follows: 
For each two-qubit gate, the single-qubit gates located immediately before or immediately after form a subblock with that gate. 
Instead of circulating ions after each gate-layer ends during execution, the proposed scheduling gives priority to this subblock execution, circulating the ions only when necessary for the next set of two-qubit gates.
In doing so, the proposed scheduling can minimize unnecessary circulation of ions during the execution.

\subsection{Basic Methodology and Principle of Proposed Scheduling} \label{propsched}

\begin{figure} [h] 
  \centerline {
  \includegraphics [width=1\columnwidth] {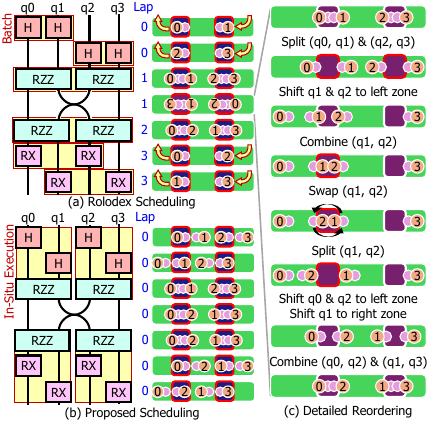} }
  \caption {
    Detailed execution scheduling comparison of 4-qubit single-layer QAOA circuits for the 4-node 2-regular target graph.
    (a) shows the execution of the existing rolodex scheduling \cite{decross2024computational}, 
    and (b) shows the execution of the proposed scheduling.
    For ease of understanding, it is considered for the racetrack electrode with two bottom gate zones and top reordering zones, respectively.
    For (a), the top zones of the racetrack electrode are utilized for ion rearrangement (it can be confirmed from the flipped numbering of ions), but for (b), both ion rearrangement and gate operations are processed using only the bottom zones.
    (c) shows a detailed ion rearrangement process on the top zones for two-qubit operations.     
  } 
  \label{f5} 
\end{figure}

The comparison between the rolodex scheduling and the proposed scheduling is illustrated through an example of a single-layer QAOA circuit for a 4-node 2-regular target graph, as depicted in the \cref{f5}.
For the intuitive explanation, assume we have a racetrack architecture with two gate operation zones.  
As shown in the \cref{f5} (a), rolodex execution computes the Hadamard gates applied to each qubit by grouping them into batches. 
Hadamard gates on qubits q0 and q1 are executed first.
Then, after q0 and q1 are pushed out in the Rolodex style, Hadamard gates on qubits q2 and q3 could be executed.
Subsequently, the two RZZ gates in the next layer are performed after the ions complete the circulation of the racetrack. 
The other two RZZs require a reordering of ions, where the qubit indexing can be changed using the top zone after the half track circulation.
Remaining operations are executed following the same principle, resulting in 3 laps to execute QAOA programs.

On the other hand, the proposed scheduling takes advantage of the opportunity to process single-qubit gates together before and after the two-qubit gate, rather than circulating the ions.
As depicted in \cref{f5} (b), Hadamard gates and their subsequent RZZ gates are identified as in-situ executable sub-blocks. 
Initially, the Hadamard gates applied to even-indexed qubits (i.e., left-side of ion pairs in the Yb-Ba-Ba-Yb crystal structure formed by two-qubit operations) are executed. 
A slight clockwise shift of ions (instead of circulating them in the rolodex style) allows the subsequent execution of Hadamard gates on odd-indexed qubits (i.e., right-side of ion pairs). 
Then, each ion pair is combined to perform the RZZ operation in place.  
As shown in sub-blocks of 1 RZZ and 2 RXs, it is straightforward to understand that these RXs immediately following the RZZ gates could be executed in place, like two H gates. 
The ion realignment operation for subsequent RZZ operations is not executed in the top zone but in the bottom zone, unlike the rolodex scheduling.
In doing so, the proposed scheduling can execute the QAOA program without any full ion's circulation.

\subsection{Implementing In-Place Executable Block-Aware Schedules} \label{lewis}

This section describes the algorithm for implementing the proposed scheduling.  
The input is a quantum circuit represented as a directed acyclic graph (DAG) \( C = (V, E) \), 
where \( V = \{ g_i \} \) is the set of 1-qubit and 2-qubit native gate operations, and  
\( E = \{ (g_i \rightarrow g_j) \} \) is the set of directed edges representing gate dependencies, such that a gate \( g_i \) should be executed before \( g_j \) if they share any qubit.  
As shown in \cref{a1}, this algorithm iteratively identifies each 2-qubit gate layer \( L_k \subseteq V \) according to qubit indices,  
where gates in each \( L_k \) are composed of mutually independent and identical types of 2-qubit gates; thereby, they can be executed in parallel.
The number of sub-blocks in \( L_k \) is determined by the architectural configuration.
For the case of H2 architecture, the value is 4.  
\( \mathcal{B}_k = \{ B_i^{(k)} \} \) is a set of subblocks in \( k \)-th layer. 

For each two-qubit gate \( g_{2q}^{(k)} \in L_k \), the algorithm identifies its adjacent 1-qubit gates (i.e., predecessors and successors), denoted \( g_{1q}^{(pre)} \) and \( g_{1q}^{(post)} \), respectively.  
Then, these are merged to form a sub-block: \( B_i^{(k)} = \{ g_{1q}^{(pre)},\; g_{2q}^{(k)},\; g_{1q}^{(post)} \} \).  
This block grouping and scheduling process is recursively applied to the subsequent layers \( L_{k+1}, L_{k+2}, \dots \) until the final layer is processed.
Within each scheduled sub-block \( B_i = \{ g_{1q}^{(pre)},\; g_{2q},\; g_{1q}^{(post)} \} \), execution proceeds as follows:  
The ion ordering is initialized such that each qubit pair \( (q_{2i}, q_{2i+1}) \) aligns with the physical Yb-Ba and Ba-Yb ion chain structure.  
If \( g_{1q}^{(pre)} \neq \emptyset \), single-qubit gates are first applied to the left qubits \( q_{2i} \), followed by a minimal ion shift to bring the right qubits \( q_{2i+1} \) into the gate zone, after which the remaining single-qubit gates are executed.  
The 2-qubit gate \( g_{2q} \) is then applied in place without global ion circulation.  
Finally, the post 1-qubit gates \( g_{1q}^{(post)} \) are executed in the same manner.

The proposed algorithm exhibits linear complexity proportional to the total number of gates in the program.
When sweeping the DAG form of the circuit once, each time a 2-qubit gate is encountered, the algorithm can merge that gate with its 1-qubit predecessor and successor gates using pre-stored pointers, without introducing any extra computational overhead.
Thus, if the circuit contains $m$ gates, the complexity of the proposed algorithm is $O(m)$.
For the case of single-layer path-graph QAOA circuits, our evaluation on systems equipped with Intel Jasper Lake N5095 processors and 32GB DDR4 memory shows that the algorithm takes 21.82 ms for a 32-qubit circuit and 27 seconds for a 1-million-qubit circuit.

\algnewcommand\algorithmicinput{\textbf{Input:}}
\algnewcommand\algorithmicoutput{\textbf{Output:}}
\algnewcommand\Input{\item[\algorithmicinput]}
\algnewcommand\Output{\item[\algorithmicoutput]}
\begin{algorithm}[h]
\caption{In-Place Executable Block-Aware Scheduling}
\label{a1}
\tiny
\begin{spacing}{0.91}
\begin{algorithmic}[]
\Input DAG-based quantum circuit \( C = (V, E) \)
\Output Optimized block-aware scheduled circuit

\State Initialize layer index \( k \leftarrow 1 \)
\While{there exist unscheduled 2Q gates}
    \State \( L_k \leftarrow \) maximal set of parallel 2Q gates of same type in \( V \)
    \State Sort \( L_k \) by qubit indices
    \State \( \mathcal{B}_k \leftarrow \emptyset \)
    \For{\( g_{2q}^{(k)} \in L_k \)}
        \State \( g_{1q}^{(pre)} \leftarrow \) 1Q predecessors of \( g_{2q}^{(k)} \)
        \State \( g_{1q}^{(post)} \leftarrow \) 1Q successors of \( g_{2q}^{(k)} \)
        \State \( B_i^{(k)} \leftarrow \{ g_{1q}^{(pre)},\; g_{2q}^{(k)},\; g_{1q}^{(post)} \} \)
        \State \( \mathcal{B}_k \leftarrow \mathcal{B}_k \cup \{ B_i^{(k)} \} \)
        \If{\( |\mathcal{B}_k| = 4 \)} \textbf{ break } \EndIf
    \EndFor
    \State Schedule all \( B_i^{(k)} \in \mathcal{B}_k \) in parallel
    \State Remove all gates in \( \mathcal{B}_k \) from \( V \)
    \State \( k \leftarrow k + 1 \)
\EndWhile

\ForAll{\( B_i = \{ g_{1q}^{(pre)},\; g_{2q},\; g_{1q}^{(post)} \} \)}
    \State Initialize ion layout: \( (q_{2i}, q_{2i+1}) \) follow Yb-Ba, Ba-Yb order
    \If{\( g_{1q}^{(pre)} \neq \emptyset \)}
        \State Apply to \( q_{2i} \); shift \( q_{2i+1} \); apply to \( q_{2i+1} \)
    \EndIf
    \State Apply \( g_{2q} \)
    \If{\( g_{1q}^{(post)} \neq \emptyset \)}
        \State Apply to \( q_{2i} \); shift \( q_{2i+1} \); apply to \( q_{2i+1} \)
    \EndIf
\EndFor
\State \( R = V \setminus \bigcup_{k,i} B_i^{(k)} \)
\end{algorithmic}
\end{spacing}
\end{algorithm}

\subsection{Improved Runtime Scalability by The Proposed Scheduling}

\begin{figure} [h] 
  \centerline {
  \includegraphics [width=1\columnwidth] {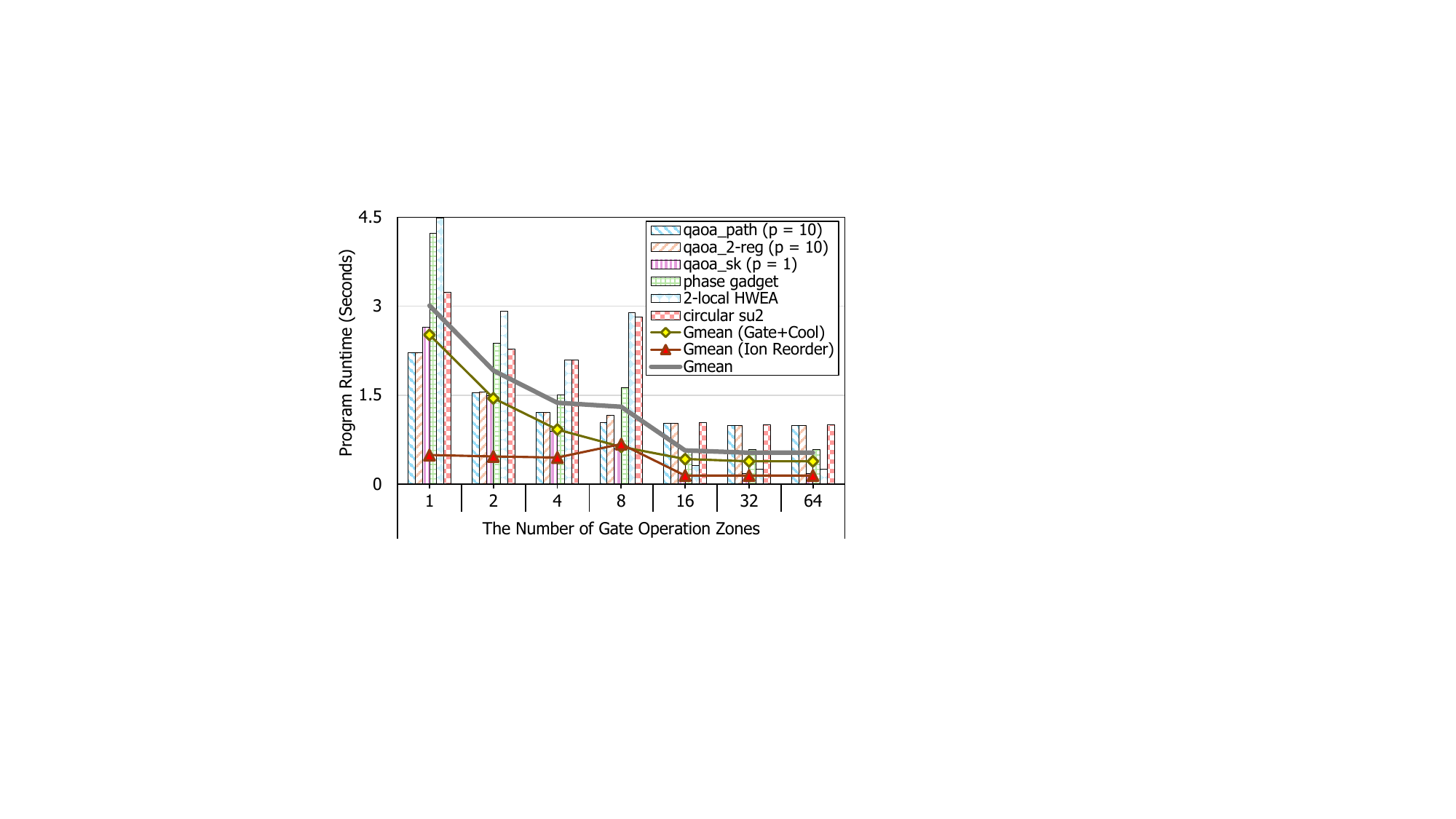} }
  \caption {
    Runtime analysis by the proposed scheduling according to the gate zone counts in the racetrack architecture using 32-qubit QAOA (for graphs of path, regular-2, and Sherrington-Kirkpatrick model) and VQE (phase gadget, 2-local hardware efficient ansatz, and circular efficient SU2 ansatz) workloads. 
  } 
  \label{f11} 
\end{figure}

The contribution of the proposed scheduling goes beyond reducing ion shuttling itself. 
The proposed scheduling suppresses the shuttling overhead, allowing the racetrack architecture to continuously improve performance as operational zones scale. 

This section analyzes the execution time performance of various 32-qubit VQA algorithms under the proposed scheduling according to the number of gate operation zones, as illustrated in the \cref{f11}. 
The only difference between the two experiments in the \cref{f2} and the \cref{f11} is the scheduling policy, and they use identical decomposition.
For the case of four gate operation zones, consistent with the current H2 architecture hardware specifications, the proposed scheduling achieves an average runtime of 1.37 seconds, which provides an 11\% reduction compared to the rolodex scheduling. 
Although it is still important to provide performance improvements in current architectures, the genuine benefit of the proposed scheduling over the rolodex is that it can provide continuous performance improvement even if the number of gate operation zones is increased.
Cases for operation zones of 16 or more provide full parallelization for handling in-place executable sub-blocks to 32-qubit programs. 
Under the proposed scheduling, these 16 or more-zone scenarios attain an average runtime of 0.53 seconds, a 2.6x improvement over the four-zone configuration. 

The proposed scheduling avoids ion circulation on the entire track by using 1-dimensional runs (e.g., H1 \cite{dasu2025breaking}) for ions when the number of zones reaches the number of maximally parallelizable sub-blocks.
Our experimental observations confirm that when the number of gate zones is sufficient, internally rearranging the ions (instead of circulating them on the entire track) can provide less runtime, even if it requires the operation of two-qubit gates in all-to-all connectivity.
This shows that even if the hardware comes in electrode shapes that allow ions to circulate, there may still be workloads that are favorable to execute partially 1-dimensional.
For example, QAOA circuits with the SK model as the target graph require all-to-all ZZ-interaction gates, which may incur substantial ion-reordering overhead.
Fortunately, this cost can be mitigated using the fermionic SWAP network approach \cite{hashim2022optimized}, which allows for all-to-all RZZ computation by exchanging only neighboring ions.

Unlike the rolodex scheduling, the proposed scheduling maintains runtime stability even when executed on systems with 16 or more gate zones. 
This means that running the identical program on a larger architecture may not degrade performance. 
Such scalability is important from a design perspective, as future versions of processors will be required to run larger-scale programs.
However, they should not offer worse performance than their previous platforms for running smaller programs, which may still be in demand for execution.

\begin{tcolorbox}[colback=green!10, colframe=green!0, boxrule=0pt, left=0.1mm, right=0.1mm, top=0.1mm, bottom=0.1mm]
\textbf{Section Summary:}
The proposed scheduling prioritizes sub-blocks composed of concurrently runnable 2Q gates and their adjacent 1Q gates, enabling scalable parallelism by suppressing the frequent ion-circulations of the entire track.
\end{tcolorbox}

\section{Logical Qubit Preparation on Racetrack Arch.}

Despite recent achievement of quantum advantage via certified randomness \cite{liu2025certificate} without error correction, the racetrack architecture would need to consider efficient preparation of error-correction (or detection) protocols to support future fault-tolerant quantum applications.
Logical qubit preparation circuits often exhibit computational properties that require identical groupings of CXs across a wide range of qubits.
Thus, it can be advantageous to increase the number of gate zones beyond the current specification to run more simultaneous gates.
Our observations focus on investigating the costs of preparing logic qubits as the architecture expands.
Our evaluations show that the most efficient specification for preparing eight logical qubits based on the [[7, 1, 3]] Steane code is 24 gate zones.
Assuming the use of a 7-qubit code, the racetrack can prepare up to eight logical qubits, which classical computers can easily simulate. 
In this context, to demonstrate the quantum advantage of the racetrack architecture for fault-tolerant algorithms, one can expect the hardware specification to be much more extended than today's, and the exploration of large-scale electrode structures will continue to be worthwhile.

\subsection{Scheduling Steane Code Circuit on Racetrack Architecture}

In this section, \([[7,1,3]]\) Steane code is considered, which is demonstrated on many quantum computers due to its simple measurement protocol and transversality of the Clifford operations \cite{ryan2021realization, ryananderson2022, paetznick2024, bluvstein2023, metodi2005quantum}.
Given that the racetrack architecture can load up to 56 physical qubits \cite{QuantinuumH2Datasheet}, it is capable of preparing up to 8 logical qubits using this code. 
Exploiting the characteristic that the CX gates within the Steane code circuit are commutable with each other allows the encoding to be executed in four gate depths (exactly, 13-depth in terms of the native gate operation) on the racetrack architectures \cite{metodi2005quantum}.

\begin{figure} [h] 
  \centerline {
  \includegraphics [width=1\columnwidth] {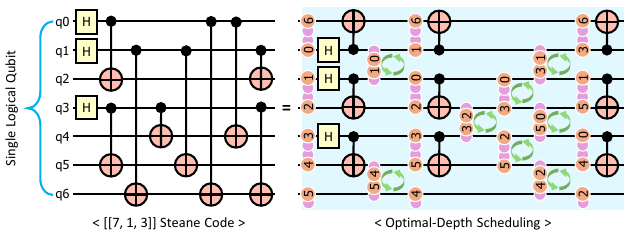} }
  \caption {
    An example of the logical qubit preparation circuit using [[7,1,3]] Steane code (left) and the proposed design with efficient shuttling scheduling for racetrack architecture (right). 
  } 
  \label{f6} 
\end{figure}

\cref{f6} shows a possible example of the proposed scheduling of preparing a single logical qubit by a single depth of Hadamard gates and three depths of CXs.
By using the proposed scheduling as discussed in \cref{lewis}, [[7, 1, 3]] Steane's code can be implemented with only four gate-depth, regardless of the number of logical qubits required.
In order to maximize gate zone utilization, it is advantageous to initially align the qubits to which the initial Hadamard gates are applied so that they are located in different pairs of ions.
Due to this, \( q_0 \), \( q_1 \), and \( q_3 \) serve as the control qubits of CX gates, while \( q_2 \), \( q_4 \), \( q_5 \), and \( q_6 \) serve as the target qubits.  
Since each logical qubit preparation circuit is separable from the others, their native operation cycle required to prepare \( n \) logical qubits in a racetrack architecture with \( k \) gate zones is given by \( 13 \cdot \left\lfloor \frac{3n}{k} \right\rfloor \).

\subsection{Ion-Reordering Detail for Preparing Fault-Tolerant Qubits} \label{reordering}

This section describes how to efficiently generate the detailed gate scheduling from the given  [[7,1,3]] Steane code circuit (on the left of \cref{f6}) to the translated circuit tailored for the racetrack (on the right of \cref{f6}).
Since the latest automated circuit compiler for the H2-1 processor \cite{decross2024computational} is currently not publicly accessible, we implement our own routing algorithm to maximize gate zone utilization. 
The primary consideration is to analyze the gate application relationship of the qubits.
To maximize gate zone utilization, gates of the same kind should be arranged as parallel as possible.
Furthermore, to minimize the ion reordering overhead, qubit pairs that are applied to 2-qubit gates frequently should be disposed close to each other. 

Therefore, the qubits of the circuit are aligned to simultaneously run the first three Hadamard gates and their following three CX gates.
The three qubits with Hadamard gates are applied with three CX gates, where they serve as control qubits for all CXs.
Nine CX gates can be processed in parallel with three depths.
Thus, one possible initial arrangement of qubits can be expressed as \((\overrightarrow 6, \overleftarrow 0), (\overrightarrow 1, \overleftarrow 2), (\overrightarrow 3, \overleftarrow 4), \overleftarrow 5\), where parentheses indicate ions coupled according to the Yb-Ba-Ba-Yb crystal structure, and the arrows above the numbers indicate the orientation of the ion pairs, with the right arrow indicating Yb-Ba and the left arrow indicating Ba-Yb. 
Some qubits (\(q_2\), \(q_4\), and \(q_5\)) serve only as the target qubit of the CX gate in this example, with only two total gates applied. 
Therefore, they are placed without forming an ion pair once each during the rearrangement process. 
The remaining four qubits (\(q_0\), \(q_1\), \(q_3\), and \(q_6\)) always participate in 3 rounds of two-qubit operations by forming ion pairs in their executions.

In the second depth of the 2-qubit gate execution step, one possible operation of another 3 CXs is processed from the pairs \((\overrightarrow 6, \overleftarrow 1)\), \((\overrightarrow 0, \overleftarrow 2)\), and \((\overrightarrow 3, \overleftarrow 5)\), and by performing only two ion-pair SWAP operations in the first gate layer, these three CX gates can be executed in parallel. 
Similarly, following the same principle, the remaining three CX gates can be executed in parallel in the third depth of the 2-qubit gate execution step.

\subsection{Estimating Logical Qubit Preparation Time by \# of Zones} \label{logicalprep}

\begin{figure} [h] 
  \centerline {
  \includegraphics [width=\columnwidth] {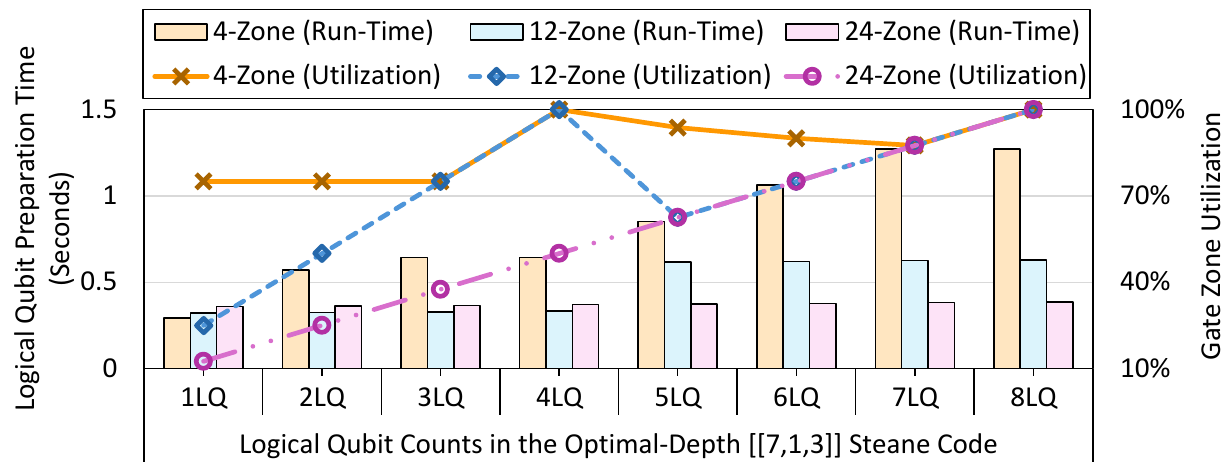} }
  \caption {
    Comparative analysis of the logical qubit preparation time and the gate zone utilization for the proposed [[7,1,3]] Steane code circuit across different numbers of gate operational zones (4, 12, and 24) as the logical qubit (LQ) increases.
  } 
  \label{f7} 
\end{figure}

This section compares the runtime and gate operation zone's utilization required for logical qubit preparation across different numbers of gate zones (4, 12, and 24), as shown in \cref{f7}.  
Note that the Steane code circuit for single logical qubit preparation has three parallel gate operations.
Therefore, the minimum architectural specification for running eight logical qubits without a circulation of ions is 24 gate zones.
Since the depth of the [[7,1,3]] Steane code circuit remains constant regardless of the qubit count, having more gate operation zones leads directly to shorter logical qubit preparation times. 
Insufficient gate operation zones could force parallel gates to be executed sequentially, increasing runtime.
As shown in \cref{f7}, as the number of logical qubits increases, we can observe steep increases in runtime in logical qubit preparation when there is a decrease in zone utilization in certain intervals.

The reason is that insufficient zone counts in the hardware require additional ion circulation in the logical qubit preparation, resulting in a sharp increase in runtime.
For example, in the 4-zone scenario, which reflects the current hardware specification of H2 architecture, preparing two logical qubits takes 96\% longer than preparing a single logical qubit. 
Similarly, in the 12-zone case, the logical qubit preparation time increases by 85\% when increasing from four to five logical qubits. 
The 24-zone configuration, which accommodates all parallel gates in a single cycle, reduces logical qubit preparation time by 18\% (on geometric mean) compared to the 12-zone, and by 51\% compared to the 4-zone.  
For the scenario with 8 logical qubits (i.e., the maximum load of qubits by H2 architecture), the 24-zone achieves runtime reductions of 38.4\% compared to the 12-zone and 68.0\% compared to the 4-zone, respectively.

Increasing the number of gate operation zones lengthens the time required for one full lap around the racetrack, revealing slight runtime increases when preparing a small number of logical qubits.  
For example, the preparation of one logical qubit in the 24-zone configuration takes 12\% longer than in the 12-zone case, and 23\% longer than in the 4-zone case. 
Nevertheless, the runtime reduction in scenarios involving a larger number of logical qubits outweighs these increases. 
For cases of preparing eight logical qubits, 12-zone is 1.9$\times$ faster than 4-zone, and 24-zone is 3.3$\times$ faster than 4-zone.
This is because, although increasing the gate operation zones to 24 lengthens the ion travel time per lap, the number of required circulations remains constant irrespective of the number of logical qubits to be prepared.
From the perspective of fault-tolerant scenarios, increasing the number of zones up to the number of simultaneously executable gates could be beneficial.  

\begin{tcolorbox}[colback=green!10, colframe=green!0, boxrule=0pt, left=0.1mm, right=0.1mm, top=0.1mm, bottom=0.1mm]
\textbf{Section Summary:}
We explore efficient 2Q gate scheduling for ECCs in racetrack processors with a case study of the Steane code.
Our evaluations confirm that 8 logical qubit preparations can be most efficiently run in the H2 processor when zone counts are increased by 6x over the current spec.
\end{tcolorbox}

\section{Introducing Shortcut in Racetrack Electrode} \label{shortcut}

This section discusses hardware modification ideas that can improve the execution efficiency of quantum programs on the racetrack architecture.
In the racetrack architecture, each qubit is circulated in a single row, and reordering of ions is performed by exchanging positions among neighboring ions.
For existing rolodex scheduling, this re-indexing is handled by a software process called automated compilation steps \cite{decross2024computational}.
Thanks to this technique, the overhead of exchanging positions of ions for each gate layer can be efficiently minimized, but these costs are still not perfectly hidden by computational time.

From the past historical perspective of computer architecture, direct hardware modifications to handle certain tasks are often more efficient and simpler to solve problems compared to their alternative software strategies \cite{ahn2005scatter}.
We can consider a scenario in which shorter paths are placed in the middle of the racetrack.
These shortcuts could provide two main advantages.

First, if we run a program that is much smaller than the number of maximum zones in the racetrack architecture, we can leverage shorter tracks as the ion's main circulation path to reduce the ion shuttling cost.
In doing so, even if the scale of the racetrack architecture expands in the future, it does not degrade the runtime performance of running small-scale quantum programs.
Second, when logical gates immediately after preparing logical qubits (encoding) are required, their re-indexing can naturally be achieved without ion split/combine.

\begin{figure} [h] 
  \centerline {
  \includegraphics [width=\columnwidth] {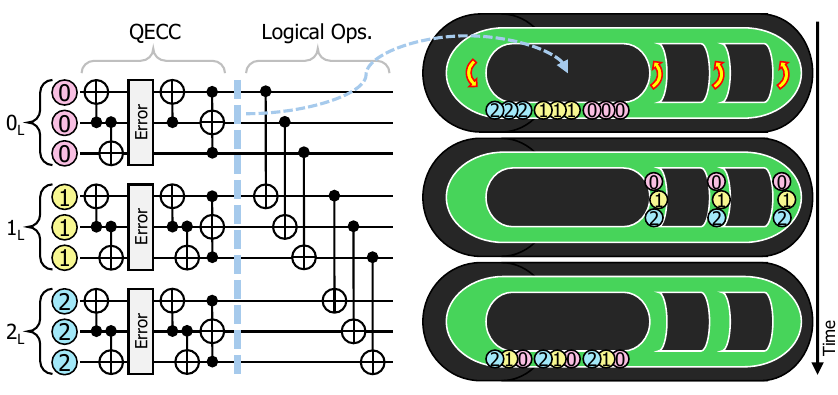} }
  \caption {
    In the left half of the figure, a circuit shows an example with three logical qubits (0$_{L}$, 1$_{L}$, 2$_{L}$), each composed of three physical qubits (000 or 111 or 222), respectively.
    In the right half of the figure, ions on the racetrack electrodes with a timeline show an example of the re-indexing process of the qubit sequence by exploiting shorter paths in a hardware-modified racetrack architecture.
    Each of the three physical qubits is prepared as 1 logical qubit using 3-qubit bit-flip ECC.
  } 
  \label{f13} 
\end{figure}

As shown in the \cref{f13}, the process of preparing a logical qubit requires gate operations between physical qubits that make up the same logical qubit, while subsequent fault-tolerant logical transversal operations \cite{wang2024optimizing, duckering2020virtualized} require gate operations between physical qubits that make up different logical qubits.
Existing rolodex scheduling has handled this through automated compilation steps, but introducing shorter paths achieves re-indexing in the next ion circulation cycle without exploiting ion reordering zones.
This is important for efficient logical program execution in the age of fault-tolerant quantum computing.
Rolodex scheduling can be viewed as a software-wise solution to ion realignment relying on optimizers by automated compilers.
We observe that the ion's reordering could be realized by the hardware without relying on the automated compiler, by a simple modification of the electrode.

Designing QCCDs to allow ions to travel various trajectories has long been discussed \cite{kielpinski2002architecture}, and recently proposed array-based and modular chip connection-based technologies would increase the reality of the shortcut \cite{brown2016co, akhtar2023high}.
Additionally, Quantinuum's latest architecture, Helios \cite{niroula2025realizationquantumstreamingalgorithm, ransford2025helios98qubittrappedionquantum, liu2025certifiedrandomnessamplificationdynamically}, adopts a lollipop-shaped electrode with a twisted ion-transport path.
This twisted track allows a smaller size of ion circulation, potentially enabling the provision of a similar shortcut effect.

\begin{tcolorbox}[colback=green!10, colframe=green!0, boxrule=0pt, left=0.1mm, right=0.1mm, top=0.1mm, bottom=0.1mm]
\textbf{Section Summary:}
We introduce a shortcut that allows ions to smaller circulate inside the racetrack, which has 2 advantages:
(i) Even if the racetrack becomes larger, it will not degrade the execution efficiency of small-scale quantum programs.
(ii) For FTQC programs, the ion rearrangement required for the logical operations after the logical qubit preparation can be achieved without relying on the gate zone.
\end{tcolorbox}

\section{Experimental Methodology}

\subsection{Modeling and Emulating Racetrack Trapped-Ion Processor}

We developed an in-house runtime emulator modeled for the Quantinuum H2-1 processor \cite{QuantinuumH2Datasheet}. 
Parameters for the emulation are listed in \cref{t0}.  
The distance between zones is 750 $\mu m$.
For scalability experiments, the change in the number of gate zones is implemented by extending or contracting the straight path of the track while maintaining this inter-zone gap.

Unfortunately, real systems, emulators, and simulations are not sufficient for performing the detailed runtime and scalability analyses required by our work \cite{QuantinuumH2Datasheet},  due to some difficulties as described below.
(i) Although the racetrack processor is accessible via cloud services, qubit routing-applied assembly-level programs and the state-of-the-art automated compiler are not available to users \cite{decross2024computational}. 
(ii) Their execution output by quantum circuits on the cloud does not report detailed runtime information.  
(iii) The existing system emulators provide infidelity estimates reflecting real hardware specifications for quantum program results, but they are not intended for runtime emulation.
(iv) Current emulators do not support architectural modifications, such as editing the number of operation zones.

\begin{table}[h]
\centering
\tiny
\renewcommand{\arraystretch}{0.82}
\renewcommand{\tabcolsep}{1mm}
\caption{Hardware Parameters for Ractrack Architecture}
\label{t0}
\begin{tabular}{|l|c|}
\hline
\textbf{Device} & \textbf{Values} \\ \hline
Zone-to-Zone Distance & 750 $\mu m$ \cite{moses2023race} \\ 
\hline
\textbf{State Perparation and Readout} & \textbf{Values} \\ \hline
Qubit initialization & 17 $ms$ \cite{moses2023race} \\ \hline
Qubit measurement (Default, high-fidelity readout protocol) & 120 $\mu s$ \cite{pino2021demonstration} \\ \hline
\hline
\textbf{Gate Operations} & \textbf{Values} \\ \hline
Single-qubit rotation ($\pi/2$) gate & 5 $\mu s$ \cite{pino2021demonstration} \\ \hline
(Native) Two-qubit gate & 25 $\mu s$ \cite{pino2021demonstration} \\ \hline
\hline
\textbf{Cooling} & \textbf{Values} \\ \hline
Cooling stage 1 (Doppler) & 550 $\mu s$ \cite{pino2021demonstration} \\ \hline
Cooling stage 2 (Axial and Radial SB) & 850 $\mu s$ \cite{pino2021demonstration} \\ \hline
Cooling stage 3 (Axial SB) & 650 $\mu s$ \cite{pino2021demonstration} \\ \hline
\hline
\textbf{Shuttling} & \textbf{Values} \\ \hline
Trap Speed at straight track & 2.65 $\mu m/\mu s$ \cite{kielpinski2002architecture, decross2024computational} \\ \hline
Moving between neighboring zones & 375 $\mu s$ \cite{kielpinski2002architecture, decross2024computational} \\ \hline
Time for 1 lap of the 4-zone racetrack & $\approx$ 6.2 $ms$ \cite{kielpinski2002architecture, decross2024computational, balensiefer2005evaluation} \\ \hline
\hline
\textbf{Ion Shift/SWAP/Split} & \textbf{Values} \\ \hline
Split or Combine & 128 $\mu s$ \cite{pino2021demonstration} \\ \hline
Intra-zone shift & 58 $\mu s$ \cite{pino2021demonstration} \\ \hline
Inter-zone shift & 283 $\mu s$ \cite{pino2021demonstration} \\ \hline
Swap & 200 $\mu s$ \cite{pino2021demonstration} \\ \hline
Ion exchange between adjacent pairs (as shown in the \cref{f5} (c)) & 1,053 $\mu s$ \cite{pino2021demonstration} \\ \hline
\hline
\textbf{Infidelity} & \textbf{Values} ($10^{-4}$) \\ \hline
1Q randomized benchmarking (RB) & 0.25 \cite{moses2023race} \\ \hline
1Q leakage & 0.04 \cite{moses2023race} \\ \hline
2Q parameterized randomized benchmarking ($\pi/2$) & 2.0 \cite{moses2023race} \\ \hline
2Q leakage & 3.9 \cite{moses2023race} \\ \hline
Transport 1Q randomized benchmarking & 2.2 \cite{moses2023race} \\ \hline
State preparation and readout (SPAM) & 16 \cite{moses2023race} \\ \hline
\hline
\textbf{Decoherence time} & \textbf{Values} \\ \hline
T1 & 100 seconds \cite{QuantinuumH2Datasheet, wang2021single} \\ \hline
\end{tabular}
\end{table}

\subsection{Runtime Factor Detail for Racetrack Processor Modeling} \label{factor}

Given the input programs, the emulator estimates the total runtime by calculating and accumulating the following factors.
The runtime of the ion-reordering is estimated as the sum of the contributions from the ion permutation and the shuttling.

The loading of the ions is already implemented in the emulator but is excluded from the evaluation, as captured ions can remain stable for tens of minutes even in the presence of background gas collisions \cite{moses2023race}.
We assume that qubits are already loaded and initially aligned at the start of the program.

\textbf{State Preparation and Readout}: 
The emulator first groups the loaded qubits into batches according to the number of gate operation zones and initializes their states to \( |0\rangle ^{\otimes n} \).
It takes $(\textrm{the number of batches})$ $\times$ 17 $ms$ to initialize qubits \cite{moses2023race}.
Moreover, we adopt a high-fidelity readout protocol \cite{pino2021demonstration}.
It takes $(\textrm{number of batches})$ $\times$ 120 $\mu s$ to measure qubits \cite{moses2023race}.

\textbf{Gate Operation}:
It is assumed that a single-qubit rotating gate operation consumes $5\mu s$.
It is assumed that the 2-qubit native gate operation consumes $25\mu s$.
For example, the gate application time required to realize a single CX gate consumes a total of 40 $\mu s$ (except for ion reordering and cooling time).

\textbf{Cooling}: 
The ion cooling process, consisting of three stages, should be applied after each gate operation. 
We assume single-qubit operations and coupling costs 2,055 $\mu s$ per batch, and 2-qubit operations and coupling costs 2,075 $\mu s$ per batch \cite{pino2021demonstration}.

\textbf{Shuttling Speed \& 1-Lap Time}: 
From the fact that the distance between two neighboring gate operational zones is 750 $\mu$m \cite{moses2023race} and the time required to shift the ions between neighboring zones is 283 $\mu$s \cite{pino2021demonstration}, we estimate that the ion pair moves along the straight path of the racetrack electrode at a speed of 2.65 $\mu$m/$\mu$s.
This speed is 4.8$\times$ faster than the shuttling speed of qubits based on Rydberg atoms \cite{evered2023high}. 
It can be estimated that a faster shuttling speed is possible because the qubits in the racetrack are ionized, so the stronger attraction acts between the qubits and the traps compared to neutral atoms.
When pairs of ions travel along the curved path of the track, they would be decelerated compared to moving along the straight path, so as not to heat them and not to shorten their lifetime \cite{kielpinski2002architecture}.
Despite the effect of this deceleration, we optimistically assume that the ion-pair will take 6.2 $ms$ to circulate the entire track.
This is to conservatively estimate the latency due to the ion's circulation by the rolodex execution.

\textbf{Ion Shift/SWAP/Split}: 
The total ion reordering time (\(T_{\text{total}}\)) can be defined as \(n_{\text{split}} \cdot T_{\text{split}} + n_{\text{swap}} \cdot T_{\text{swap}} + n_{\text{inter-zone}} \cdot T_{\text{inter-zone}} + n_{\text{intra-zone}} \cdot T_{\text{intra-zone}}\), where \(n_{\text{x}}\) is the total number of the re-odering operation x, \(T_{\text{x}}\) represents the consumed time for the single re-ordering operation x.
A split operation separates an ion pair, while a combine operation merges two ions into a pair: \((\overrightarrow 0,\overleftarrow 2) \leftrightarrow \overrightarrow 0, \overleftarrow 2\). 
A swap operation means exchanging the positions of two ion pairs within the same operation zone: \((\overrightarrow{0}, \overleftarrow{2}) \rightarrow (\overrightarrow{2}, \overleftarrow{0})\). 
An inter-zone shift operation means moving an ion pair to the adjacent zones. 
An intra-zone shift operation means a single Yb-Ba or Ba-Yb passes via a zone.

\subsection{Fidelity Factor Detail for Racetrack Processor Modeling} \label{fidelityfactor}

The emulator estimates the overall fidelity (or infidelity \cite{jang2023quixote}) of hardware executions by combining contributing factors as follows.
The overall infidelity is obtained as $I_{\mathrm{total}} = 1 - F_{\mathrm{total}}$, where $F_{\mathrm{total}}$ represents the accumulated fidelity from various error sources.
$F_{\mathrm{total}}$ is obtained by
$F_{\mathrm{SPAM}} \times F_{\textrm{1Q Gate}} \times F_{\textrm{2Q Gate}} \times F_{\mathrm{transport}} \times F_{\mathrm{decoh.}}$ \cite{wang2024atomique}.
The total 1-qubit (1Q) or 2-qubit (2Q) gate fidelity for the input program can be calculated as $F_{\mathrm{Gate}} = F_{\mathrm{RB}} \times F_{\mathrm{l}}$, where $F_{\mathrm{RB}}$ denotes the randomized benchmarking fidelity and $F_{\mathrm{l}}$ the leakage fidelity.

The \textit{Transport 1Q RB} term models errors arising from ion-pair reordering and from the acceleration or deceleration of ions during shuttling.
In this work, one \textit{Transport 1Q RB} event is accumulated whenever an ion is exchanged between pairs or when an ion pair passes through a curved electrode of the racetrack.
The decoherence-induced fidelity's decay follows the exponential model $F_{\mathrm{decoh.}} = e^{-t/T_1}$, where $t$ is the emulated run-time and $T_1$ is qubit's relaxation time \cite{jang2024barber}.

This estimation tracks how much each final amplitude dissipates into others.
Programs that provide multiple answers could achieve higher effective fidelities than this assumption.

One potential concern is that the increased parallelism by the proposed method can lead to higher gate crosstalk.
To our knowledge, no gate crosstalk was reported for H2.
Instead, reset crosstalk was reported \cite{moses2023race}, much lower than 1Q RB infidelities.
Thus, we assume that gate crosstalk is negligible.

\subsection{Near-Term or Fault-Tolerant Quantum Circuit Benchmarks} \label{benchs}

The experiments in this research consider QAOA (quantum approximate optimization algorithm \cite{marwaha2021local, hashim2022optimized, jang2024recompiling}) and VQE (variational quantum eigensolver \cite{ueno2024sfq}) benchmarks as near-future quantum applications, and [[7, 1, 3]] Steane Code \cite{ryan2021realization, ryananderson2022, paetznick2024, bluvstein2023, metodi2005quantum}, Magic State Distillation \cite{bravyi2005universal, daguerre2025experimental}, and GHZ (Greenberger-Horne-Zeilinger) State circuit \cite{cruz2019efficient} as fault-tolerant applications.
Specifically, we consider single-layer QAOA circuits for Maximum-Cut with target graphs of path-shaped \cite{tannu2019ensemble}, power-law \cite{ayanzadeh2023frozenqubits}, 2-regular \cite{marwaha2021local}, and Sherrington-Kirkpatrick model \cite{farhi2022quantum} and VQE circuits of Z-phase gadget \cite{Cowtan_2020}, 2-local HWEA \cite{li2024case}, and circular-shaped efficient SU2 \cite{wang2023prepare}. 
For the Magic State Distillation program, it is considered for the 7-to-1 T-State Distillation circuit (please refer to Figure 32 on the \cite{fowler2012surface}).
The GHZ state circuit adopts a parallel chain structure in which the depth of the CX gates expands logarithmically with the number of qubits \cite{cruz2019efficient} as a baseline form, which is advantageous in terms of gate zone utilization.

Error correction codes do not apply in near-future quantum applications, and in fault-tolerant quantum applications, it is assumed that seven physical qubits constitute one logical qubit.

We also consider the encoding cost of the quantum Reed–Muller (QRM) codes \cite{barg2025geometric}.
Specifically, we focus on the case where $n = k + 2$,
where $n$ and $k$ denote the number of physical and logical qubits, respectively \cite{barg2025geometric}.
Such codes can implement a logical Hadamard gate by combining transversal Hadamard gates and ion permutation across all physical qubits.

\subsection{Further Execution Optimization: Hiding Runtime Latency} \label{gatepipe}

This section discusses a simple strategy to hide runtime latency in the racetrack architecture.
The rolodex scheduling first performs initialization to all qubits during the first ion circulation, and then performs the first gate application at the next qubit circulation \cite{decross2024computational}.
Similarly, rolodex scheduling applies the last gates to all qubits during the second-to-last qubit circulation, and then they are measured at the last qubit circulation cycle.
However, with the same principle discussed in \cref{lewis2}, initialization and measurement processes can be executed with their adjacent gates with priority over ion movements.
The proposed gate pipelining integrates qubit initialization and measurement operations within the gate application process.
When gate pipelining is applied, the initialized qubits are applied gate operations as continuously as possible rather than immediately circulating the track.
Similarly, the qubits applied (the set of) last gates are measured immediately, rather than being measured on the next cycle of the ion circulation.

Furthermore, doing as many tasks as possible before the ion circulation also helps to hide the ion circulation time itself.
For example, while processing the operation on the last gate batch for the current gate-layer, the ions for the first gate batch for the next gate-layer can cycle through the track in advance to prepare for the next gate operation to run immediately.
Our experimental evaluations confirm that this gate pipelining strategy can hide the qubit initialization time on the four-zone racetrack case by an average of 87\% for 32-qubit workloads.

\section{Experimental Results and Analyses} \label{results}

\subsection{Runtime Evaluations with Near-Term Variational Programs} \label{specific}

\begin{figure} [h] 
  \centerline {
  \includegraphics [width=1\columnwidth] {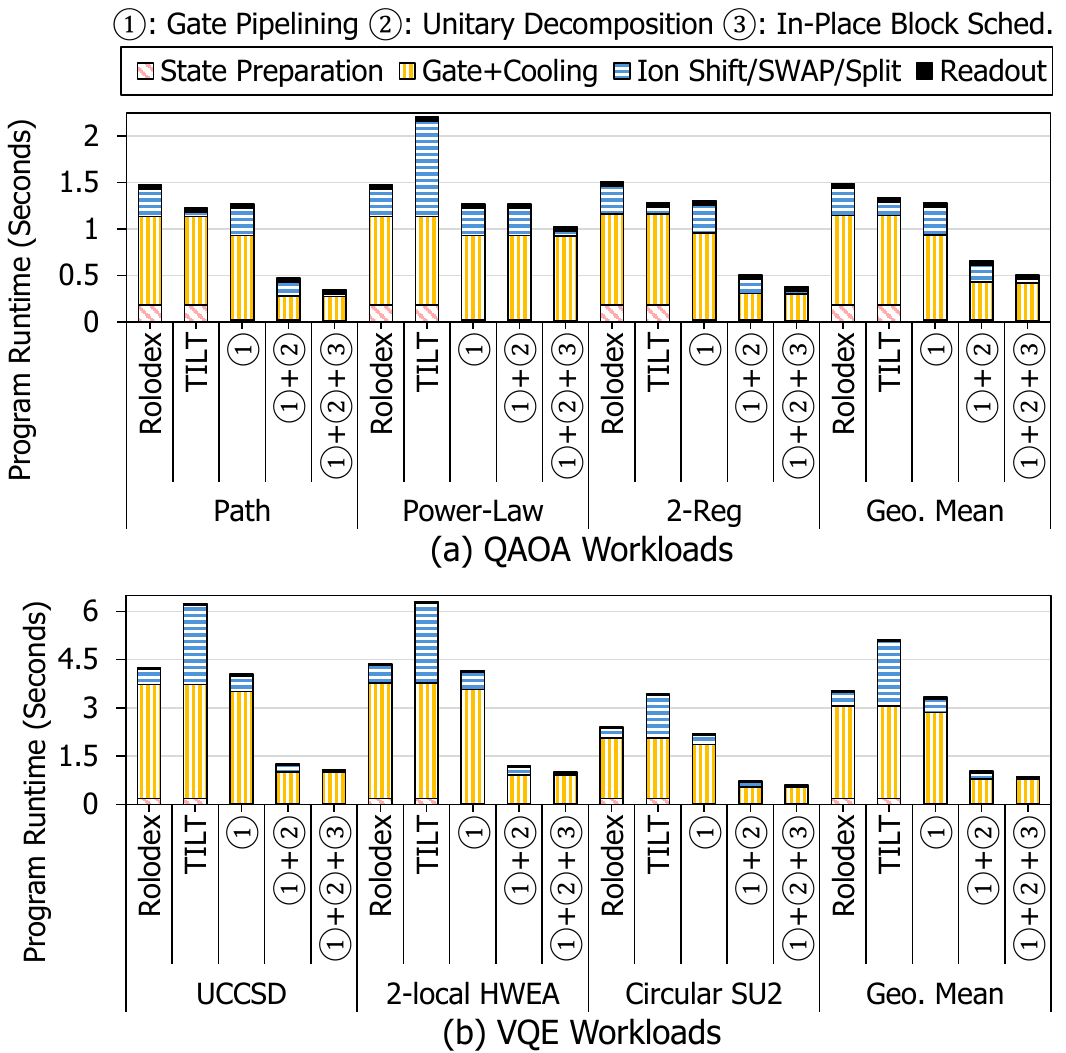} }
  \caption {
    Runtime breakdown analysis of 32-qubit near-future workloads under various scheduling strategies.
    (a) shows the runtime results for QAOA workloads across programs: Path, Power-Law, 2-Regular, and their geometric mean.
    (b) shows the runtime results for VQE workloads across programs: UCCSD, 2-Local HWEA, Circular SU2, and their geometric mean.
    Programs are evaluated under 5 scheduling strategies: Rolodex, TILT, 
    \textcircled{1} (Gate Pipelining),
    \textcircled{1}+\textcircled{2} (adds Unitary Decomposition), and \textcircled{1}+\textcircled{2}+\textcircled{3} (adds In-Place Block Sched.).
  } 
  \label{f10} 
\end{figure}

This section analyzes the runtime breakdown analysis, applying the proposed scheduling methods for 32-qubit QAOA and VQE workloads, as shown in \cref{f10}.
\textcircled{1} scheduling represents applying the Gate Pipelining strategy as discussed in \cref{gatepipe}, and provides 14\% runtime savings on average QAOA workloads and 6\% on VQE workloads, respectively, compared to the rolodex scheduling.
Thanks to the addition of the Unitary Decomposition as discussed in the \cref{decomp}, \textcircled{1}+\textcircled{2} leads to more runtime reductions. 
This combination of \textcircled{1}+\textcircled{2} achieves a 56\% decrease in runtime for QAOA and 71\% decrease for VQE workloads compared to the Rolodex. 

While ion-trap quantum computers benefit from rapid shuttling speeds compared to other atom-based systems, such as Rydberg atoms \cite{evered2023high, lin2025reuse, jang2025qubit}, ions may be constrained by a limited number of gate zones. 
As discussed in Section 3, the proposed Unitary Decomposition maximizes the utilization of these gate operation zones, thereby reducing execution times. 
Through optimized decomposition and relocation of operations to facilitate parallel execution, the \texttt{Gate+Cooling} time and the \texttt{Ion Shift/SWAP/Split} time could be concurrently minimized. 
This reduction does not stem from a decrease in the number of operations for ion re-ordering but rather from the enhanced gate-level parallelism by the program re-writing.

Further augmenting the scheduling strategy by the In-Place Block Scheduling (as discussed in \cref{lewis2}) by \textit{Plutarch} (\textcircled{1}+\textcircled{2}+\textcircled{3}) yields additional runtime savings, achieving a 66\% reduction for QAOA and 75\% for VQE relative to the baseline. 
In contrast to rolodex, which necessitates circulations of ions around the track for every gate-layer in the program, In-Place Block Scheduling prioritizes the execution of local blocks that can be performed without the entire ion-circulation. 

\textbf{Compared with H1}:
This section further discusses comparisons with the H1 \cite{dasu2025breaking}-like run.
We refer to this as TILT, but note that it differs from the hardware in the original work \cite{wu2021tilt}.
TILT intentionally utilizes only the bottom zones, executing operations in a 1-dimensional manner.
TILT moves ions from side to side along the track for computation in each gate-layer.
For an $n$-qubit program, the time required to sweep once for all qubits across the zones can be estimated as $(n-1) \times$ \textit{inter-zone shift} time \cite{wu2021tilt}.
The comparison with TILT provides two insights.
First, as shown in \cref{f10} (a), for some cases where native 2-qubit gates are placed adjacently, processing them one-dimensionally rather than circulating ions for every gate layer can effectively reduce shuttling time.
Nevertheless, TILT may face two limitations due to its inability to perform ion circulation:
(i) it forces slow ion pair permutations when two-qubit gates should be applied between distant qubits, and
(ii) it cannot exploit the pipelining advantage between gate execution and ion's reordering, resulting in an average longer overall runtime.
In contrast, the Rolodex can detour these limitations by enabling faster reordering when distant two-qubit gates are required, as some qubits can return via the upper track.
Rolodex also allows pipelining: some qubits perform gate operations in the bottom zones while others are reordered in top zones.
TILT reduces 6\% runtime on QAOA workloads, but increases 45\% on VQE workloads, compared to Rolodex.

Interestingly, as shown in \cref{f10} (a), TILT reduces the overall runtime by an average of 6\% compared to rolodex, primarily by consuming less \texttt{Ion Shift/SWAP/Split} time for the case of Path and 2-Reg.
This relative improvement arises because TILT's 1-dimensional execution can process native 2-qubit gates (here, RZZ) more efficiently when they are placed consecutively between neighboring ions, unlike Rolodex's ion circulation-priority policy for every gate-layer.

\textbf{Estimating full QAOA training runtime}:
We exploit the execution budget reported in a recent end-to-end QAOA training demonstration on the real H2-1 \cite{hao2024end}.
For 32-qubit 5-layer QAOA circuits, using 10,000 shots per epoch, their experiments achieve the approximation cut ratio \cite{jang2024recompiling} close to that of the noiseless state-vector simulation \cite{lee2025pimutation} after just two iterations \cite{hao2024end}.
According to this set-up, the estimated per-epoch circuit runtime for three target graphs in \cref{f10} using the Rolodex approach is 20.69 hours on average.
Note that the target graphs considered in this work are relatively sparse.
Thus, with modern GPU-equipped classical computing resources \cite{choquette2023nvidia} and an efficient optimizer such as COBYLA \cite{powell2007view}, the parameter update overhead is expected to be less than 10 minutes per epoch, which is negligible compared to the QPU's runtime.
Consequently, the end-to-end runtime of the QAOA applications under the rolodex approach is estimated to be 41.38 hours on average.
In contrast, the total runtime of 3 QAOA applications using the \textit{Plutarch} approach (\textcircled{1}+\textcircled{2}+\textcircled{3}) is estimated to be 14.07 hours on average.
This shows that the proposed techniques could reduce the end-to-end parameter training runtime of QAOA applications to a meaningful scale.

\subsection{Fidelity Evaluations with Near-Term Variational Programs} \label{fidelity}

\begin{figure} [h] 
  \centerline {
  \includegraphics [width=1\columnwidth] {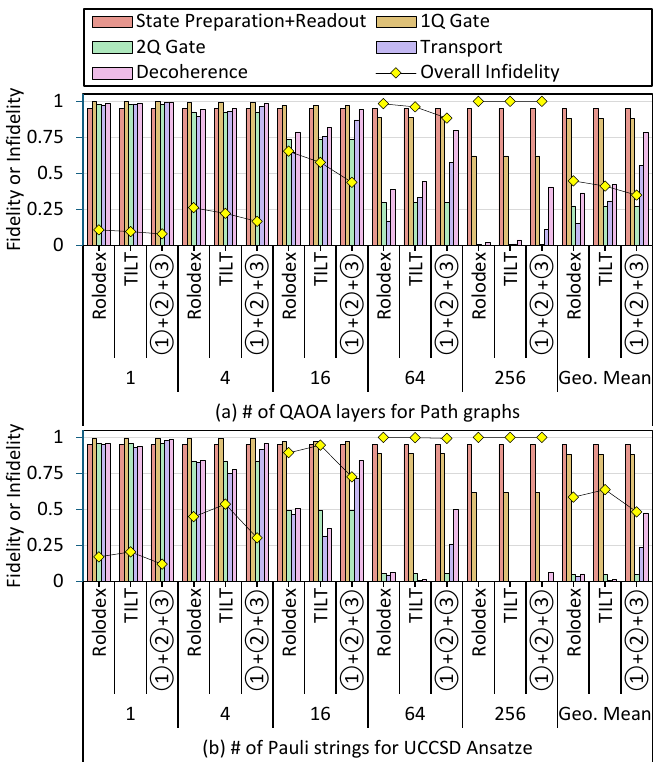} }
  \caption {
    Fidelity analysis of 32-qubit near-future workloads as circuit depth increases.
    (a) shows QAOA workloads for path-shaped graphs as the number of layers quadruples.
    (b) shows VQE workloads for the UCCSD ansatz as the number of Pauli strings quadruples.
    Programs are evaluated under 3 scheduling strategies: Rolodex, TILT, and \textcircled{1}+\textcircled{2}+\textcircled{3} (\textit{Plutarch}).
    Bars represent fidelity contributions from each component (higher is better), and lines mean the overall infidelity (lower is better).
  } 
  \label{f15} 
\end{figure}

We evaluate the fidelities or infidelities of 32-qubit near-future workloads (QAOA for path-shaped graphs and VQE for UCCSD ansatz) under various scenarios as program depth increases in \cref{f15}.
By employing extremely deep circuits (up to 256 layers or Pauli strings), we analyze the contribution of each fidelity component to the overall error accumulation.
Across most program scheduling scenarios, errors originating from two-qubit gate operations and ions' transport accumulate most rapidly as the circuit depth grows, whereas fidelity associated with single-qubit operations remains relatively robust.

For the case of H1-like run (TILT), the overall infidelity is 7.93\% lower than Rolodex on average for QAOA workloads but 8.88\% higher for VQE workloads.
This indicates that the one-dimensional execution of TILT could be restrictively effective than Rolodex when executing consecutively applied hardware-native $RZZ$ gates that constitute the cost Hamiltonian of QAOA, but not for the more general case of two-qubit gate chain structures such as the UCCSD ansatz.
In scenarios requiring consecutive native two-qubit gate operations among nearby qubits, the H1-style side-to-side ion movement could be more efficient than the full-track circulation performed at every gate layer by Rolodex.
However, since TILT deliberately prohibits ion circulation, it may incur costly reordering within the one-dimensional ion array when long-distance two-qubit interactions are frequently required, as in UCCSD workloads.

\textit{Plutarch} achieves lower overall infidelity than both baselines: 19.73\% lower than Rolodex and 19.83\% lower than TILT on average.
This improvement may primarily stem from (i) the suppression of excessive ion circulation compared to Rolodex and (ii) the use of appropriate ion circulation for efficient reordering, unlike the strict 1-dimensional execution.

When analyzing the fidelity by each contributing factor, the relative improvements by \textit{Plutarch} over the two baselines would mainly originate from differences in \texttt{Transport} and \texttt{Decoherence}.
Since \textit{Plutarch} does not alter the number of qubits or gate unitary itself in the original program, state-preparation and readout errors and gate errors accumulate nearly equivalently across all three scheduling strategies.
However, because \textit{Plutarch} circulates ions only when necessary, the transport-induced errors accumulate more slowly, and the reduced overall runtime achieved by the proposed scheduling leads to stronger robustness against decoherence error.
According to our evaluation, the average transport fidelity of \textit{Plutarch} is 4.79$\times$ and 7.58$\times$ higher than that of TILT and Rolodex, respectively, while the decoherence fidelity is 4.33$\times$ and 7.94$\times$ higher, respectively, across the 2 VQA workloads.

\subsection{Execution Time Evaluations with Fault-Tolerant Programs} \label{ft-programs}

This section discusses runtime comparison evaluations according to different scheduling strategies for fault-tolerant workloads.
All workloads are assumed to prepare logical qubits via the [[7,1,3]] Steane code.
Unlike the above evaluation of workloads in \cref{specific}, applying the scheduling techniques proposed in this work does not drastically reduce the \texttt{Gate+Cooling} time of FTQC workloads.
The reason is that the circuits of error correction codes have a configuration that repeatedly applies the identical CX gate structure to different physical qubit groups.
Because these gate configurations are already sufficiently parallelized and written in a separable manner \cite{duan2009distinguishability}, the gate operation zones are already being utilized to the maximum by the Rolodex method alone.

\begin{figure} [h] 
  \centerline {
  \includegraphics [width=1\columnwidth] {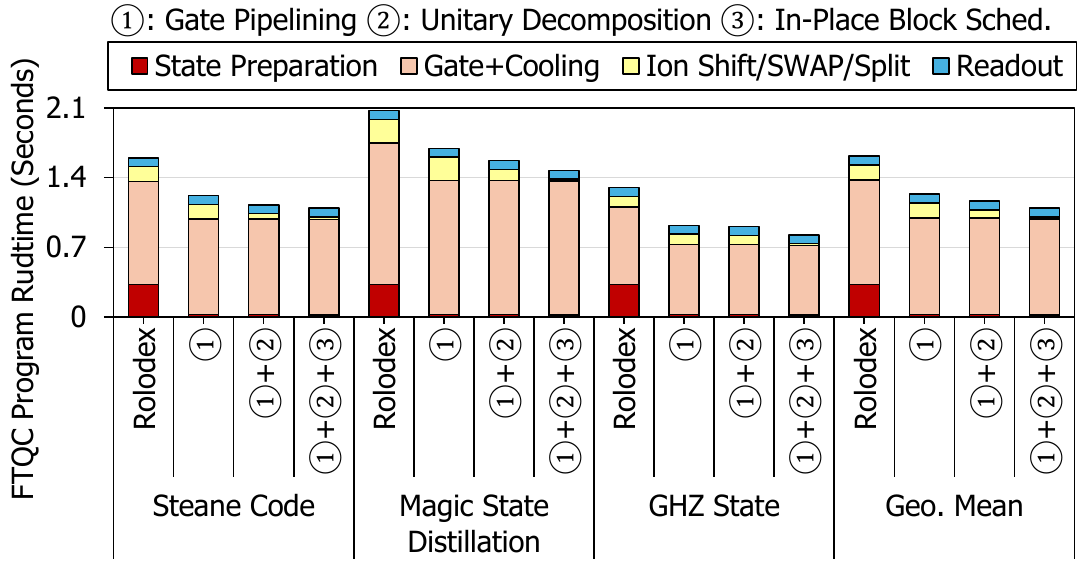} }
  \caption {
    Runtime breakdown results for 8-logical qubit (i.e., 56-physical qubit) FTQC workloads: Steane Code, Magic State Distillation, GHZ State, and their geometric mean.
    Each program is evaluated under the four scheduling strategies: Rolodex (baseline), 
    \textcircled{1} (Gate Pipelining),
    \textcircled{1}+\textcircled{2} (adds Unitary Decomposition), and \textcircled{1}+\textcircled{2}+\textcircled{3} (adds In-Place Block Sched.).
  } 
  \label{f12} 
\end{figure}

As shown in \cref{f12}, \textcircled{1} scheduling reduces the average runtime by 24\%, \textcircled{1}+\textcircled{2} scheduling reduces the average runtime by 28\%, and \textcircled{1}+\textcircled{2}+\textcircled{3} (\textit{Plutarch}) scheduling reduces the average runtime by 32\%, compared to the rolodex scheduling.
For the evaluations of variational algorithm workloads in the \cref{specific}, Unitary Decomposition contributes the most to runtime reduction.
On the other hand, in FTQC workloads, we can observe that the latency hiding effect of ion circulations by Gate Pipelining contributes the most to the runtime reduction.

Given the 2-qubit gate fidelity (99.914\% \cite{QuantinuumH2Datasheet}) of the current H2 architecture, a single recursion of ECC may not be enough to reach fault-tolerant quantum computing in a strict sense.
We estimate that at least logical error rates of $10^{-6}$ should be required for the genuine FTQC.
To obtain better logical error rates, ECC would need to be encoded multiple times, and estimating their overhead will be interesting future works.

\subsection{Electrode-Modified Evaluations for Racetrack Processors} \label{modified}

\begin{figure} [h] 
  \centerline {
  \includegraphics [width=1\columnwidth] {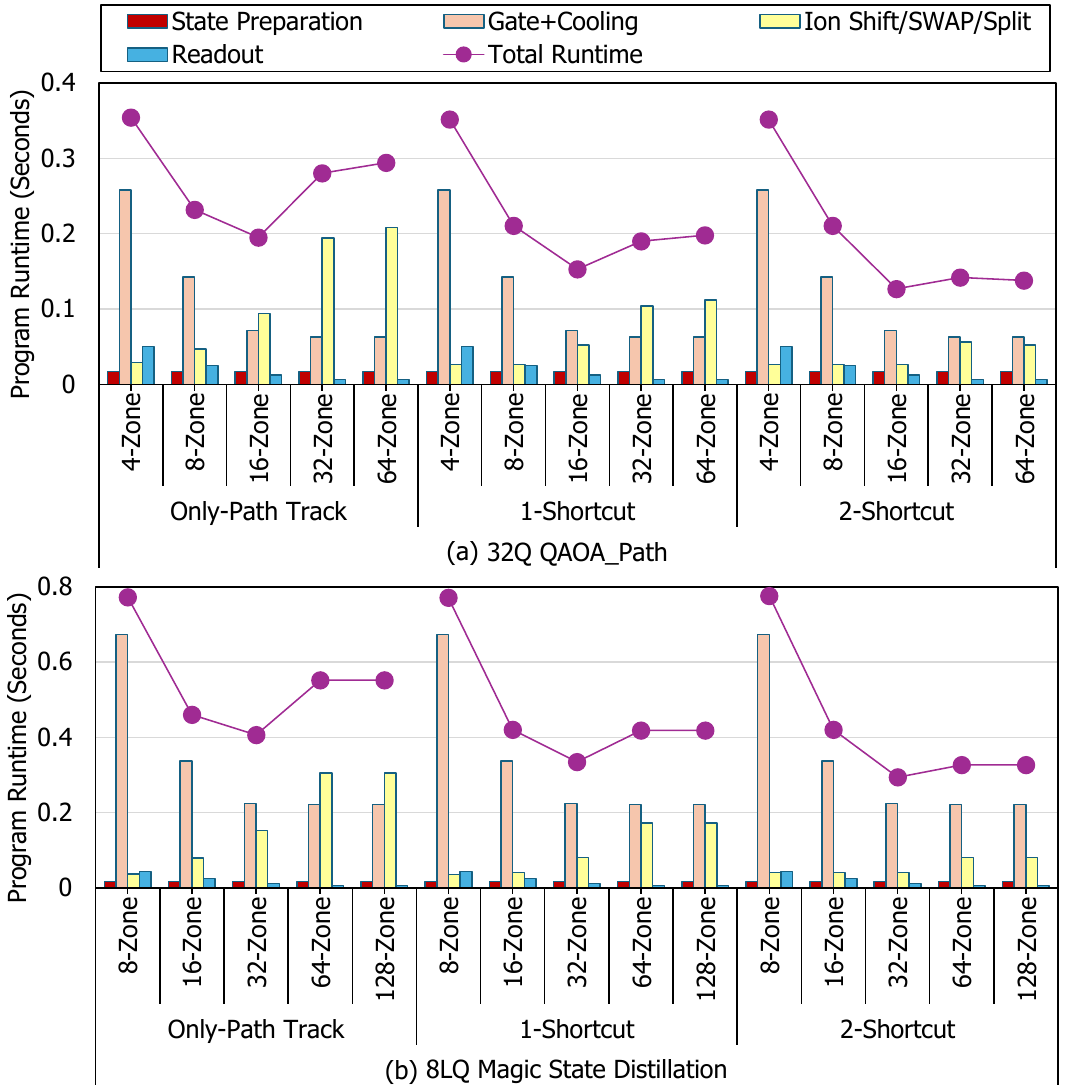} }
  \caption {
    Runtime breakdown analysis of various numbers of shortcuts (only path, a single shortcut, two shortcuts) and gate operation zones (4 to 128 zones).
    (a) shows the evaluations of the 32-qubit QAOA Path workloads, and (b) shows the evaluations of 8-logical qubit Magic T-State Distillation workloads.
  } 
  \label{f14} 
\end{figure}

This section discusses evaluations that extend the Racetrack architecture to have various operation zones and shortcuts.
For \texttt{Only-Path Track} (i.e., no shortcut), as shown in \cref{f14} (a), the 32-qubit QAOA Path workload can run most efficiently when there are 16 gate zones.
Similarly, for the \texttt{Only-Path Track}, as shown in \cref{f14} (b), the 8-logical qubit magic state distillation workload can run most efficiently when there are 32 gate zones.
The reason is that as the number of gate zones increases, the number of gates that could be simultaneously processed increases, so the \texttt{Gate+Cooling} time decreases, but the time for the ions to circulate the racetrack increases.
As noted in \cref{lewis2}, the scheduling proposed in this work suppresses ion shuttling overhead by shifting the ions via one-dimensional H1-style execution (instead of circulating ions) when the number of gate zones in the hardware is greater than the maximum number of parallel applied gates.
Nevertheless, the increase in gate zones leads to an increase in total runtime, which may degrade the performance for the identical program.

When the shortcuts are added, the ions are allowed to selectively circulate to a shorter electrode path, thereby suppressing shuttling time.
We consider a scenario in which shortcuts are added to points corresponding to exactly $\tfrac{1}{2}$ (for 1-Shortcut and 2-Shortcut) and $\tfrac{1}{4}$ (for 2-Shortcut) of the straight line of the electrode, respectively.
As shown in \cref{f14} (a), when there are 64 gate zones, the runtime performance is improved by 33\% when each has one shortcut, and runtime performance is improved by 53\% when there are two shortcuts, compared to \texttt{Only-Path Track}.
Similarly, as shown in \cref{f14} (b), when there are 64 gate zones, the runtime performance is improved by 24\% when each has one shortcut, and runtime performance is improved by 41\% when there are two shortcuts, compared to \texttt{Only-Path Track}.
Therefore, adding adequate shortcuts could provide better runtime scalability, even if the racetrack architecture becomes larger in the future.

\subsection{Efficient Shortcut Layout for Large-Scale ECC Encoding} \label{shortcut2}

\begin{figure} [h] 
  \centerline {
  \includegraphics [width=1\columnwidth] {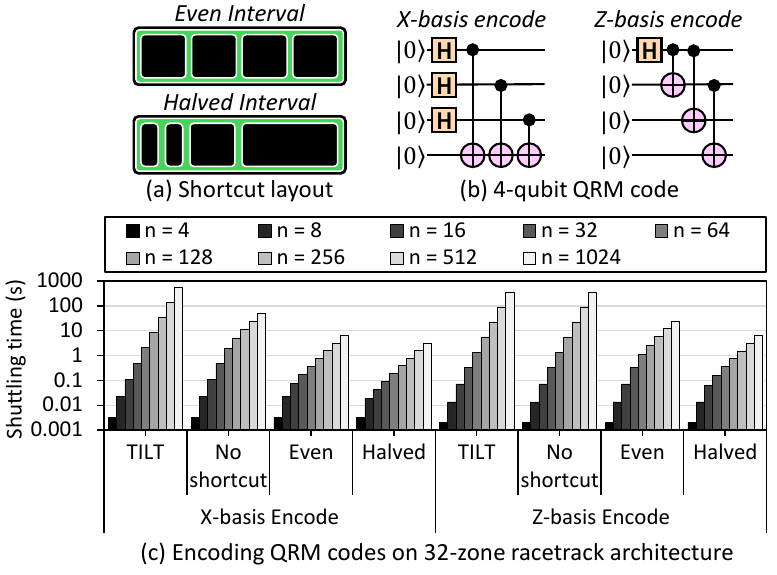} }
  \caption {
    (a) shows two shortcut layouts, where 32 operational zones are distributed across the top and bottom segments in proportion to the sub-electrode length.
    (b) shows the encoding circuits for logical qubit preparation in the X- and Z-basis using a [[4,2,2]] QRM code \cite{liu2025coniq} (when \# of physical qubits $n=4$).
    (c) shows the shuttling time analysis of the encoding under various electrodes' layouts as the number of $n$ doubles.
  } 
  \label{f16} 
\end{figure}

We evaluate the ion transport and reordering cost for efficient QRM encoding under various layouts, as shown in \cref{f16}.
The cost of rearranging ions for encoding without track circulation can increase quadratically with the number of qubits, whereas introducing some shortcuts reduces this scaling to linear.
We observe that non-uniform interval layouts, such as the \textit{Halved} configuration, can provide superior efficiency.
For instance, when identical 1-qubit gates are required for all qubits, circulating along the longest path can maximize parallelism.
In contrast, when local entangling operations are required, ions' circulations along the shortest path could enable faster return and reduce latency.
Our evaluation shows that the \textit{Halved} layout reduces shuttling cost by 48.99\% compared to \textit{Even}, 81.13\% compared to \textit{No Shortcut}, and 86.38\% compared to \textit{TILT}.
This suggests that, for future large-scale QCCD architectures \cite{jones2025architecting}, introducing several short paths can be more effective than expanding a uniform grid with equal intervals.

Specifically, for the case of encoding X-basis state, as the number of physical qubits $n$ increases, the circuit requires long-range CX operations with a maximum distance of $n-1$ qubits in the worst case.
This means that, regardless of the initial physical qubit mapping and alignment, H1-style one-dimensional execution should rely on relatively slow ion-reordering zones to complete the encoding process.
In contrast, the Rolodex (\texttt{No shortcut}) approach can transform these long-range CX operations into relatively short-range ones through ion circulation, reducing the average shuttling cost by up to 11.1$\times$ compared to TILT.
For layouts such as \textit{Even} and \textit{Halved}, the ions are allowed to return via shorter circulation paths for such near-range CX operations.
Our evaluation shows that \textit{Even} and \textit{Halved} layouts reduce shuttling costs by up to 7.84$\times$ and 15.66$\times$, respectively, compared to \texttt{No shortcut}.

For the case of encoding Z-basis states, this workload could also be recognized as efficiently preparing an $n$-qubit GHZ state.
As discussed in \cref{f13}, introducing appropriate shortcuts enables parallel CX chain execution immediately after encoding without relying on the ion reordering zones.
This is achieved by allowing ions to permute naturally through multiple electrode paths, which requires at least two circulation routes, i.e., at least one shortcut.
Due to this, in the case of Z-basis encoding, \texttt{No shortcut} layout does not show dramatic improvements over TILT.
However, the \textit{Even} and \textit{Halved} layouts can reduce shuttling costs by up to 14.72$\times$ and 58.13$\times$, respectively, compared to the No shortcut layout.

\section{Related Works}

A recent demonstration using a code switching \cite{daguerre2025experimental} for efficiently preparing high-fidelity magic states \cite{dasu2025breaking} was conducted on the real racetrack processor, leveraging the transversal property of the CSS (Calderbank–Shor–Steane) code \cite{steane2002quantum}.
Additionally, a demonstration for the fault-tolerant implementation of the Deutsch–Jozsa algorithm \cite{deutsch1992rapid} using [[4,2,2]] QRM code was conducted on the real racetrack processor \cite{singh2024fault}.

\section{Conclusion}

The electrode structures for trapped-ion quantum computing machines have the potential to expand more complexly (beyond the racetrack) in the future.
Strategies proposed in this work would be helpful for the scalable design of their systems.

\section*{Acknowledgments}
The authors thank the reviewers for their constructive comments.
The authors also thank Zichang He and Tianyi Hao for discussions on QAOA runtime estimation.
This work was funded by the National Research Foundation of Korea (NRF) under the project, “Creation of the Quantum Information Science R\&D Ecosystem Based on Human Resource” (RS-2023-00303229). 
This research was also supported by the education and training program of the Quantum Information Research Support Center, funded via the NRF by the Ministry of Science and ICT (MSIT) of the Korean government (RS-2023-NR057243).
Won Woo Ro is the corresponding author.

\bibliographystyle{IEEEtranS}
\bibliography{refs}

\end{document}